\documentclass[letterpaper]{JHEP3}

\usepackage[dvips]{graphicx}
\usepackage{epsfig,multicol,bbm}
\usepackage{amsmath}
\usepackage{appendix}

\def\lsim{\mathrel{\hbox{\rlap{\hbox{\lower4pt\hbox{$\sim$}}}\hbox{$<$}}}}

\makeatletter

\newcommand{\Rmnum}[1]{\expandafter\@slowromancap\romannumeral #1@}
\makeatother

\title{Channeling in direct dark matter detection II : \\ channeling fraction in Si and Ge crystals}

\author{Nassim Bozorgnia\\
Department of Physics and Astronomy, UCLA, 475 Portola Plaza, Los
  Angeles, CA 90095, USA\\
E-mail: \email{nassim@physics.ucla.edu}}

\author{Graciela B. Gelmini\\
Department of Physics and Astronomy, UCLA, 475 Portola Plaza, Los
  Angeles, CA 90095, USA\\
Email: \email{gelmini@physics.ucla.edu}}

\author{Paolo Gondolo\\
Department of Physics and Astronomy, University of Utah, 115 South 1400 East \#201,
  Salt Lake City, UT 84112, USA\\
  School of Physics, KIAS, Seoul 130-722, Korea\\
E-mail: \email{paolo@physics.utah.edu}}

\abstract{The channeling of the ion recoiling after a collision with a WIMP changes the ionization signal
in direct detection experiments, producing a larger signal than otherwise expected. We
give estimates of the fraction of channeled recoiling ions in Si and Ge crystals
using analytic models produced since  the 1960's and 70's  to describe channeling and blocking effects.
We used data obtained to avoid channeling in the implantation of dopants in Si crystals to test our models.}

\begin{document}

\section{Introduction}

Channeling and blocking effects in crystals refer to the orientation dependence of charged ion penetration in crystals. In the ``channeling effect" ions incident upon a crystal along symmetry axes and planes suffer a series of small-angle scatterings  that maintain them in the open ``channels"  in between the rows or planes of lattice atoms and thus penetrate much further into the crystal than in other directions.
  Channeled incident ions do not get close to lattice sites, where they would be deflected at large angles.

 The ``blocking effect"  consists in a reduction of the flux of ions originating in lattice sites along symmetry axes and planes, due to large-angle scattering with the atoms immediately in front of the originating lattice site, creating what is called a ``blocking dip" in the flux of ions exiting from a thin enough crystal as function of the exit  angle with respect to a particular symmetry axis or plane.

  Channeling and blocking effects in crystals
   are related because the non-channeled incident ions are those which suffer a close-encounter process with an atomic nucleus in the crystal, namely those which pass sufficiently close to a lattice nucleus to be deflected at a large angle. After a close-encounter collision the deflected ion acts as if it was ``emitted" from a lattice site. Channeling is many times observed as a lack of ions  (incident at a small angle $\psi$ with respect to a particular symmetry axis or plane) deflected at a large-angle which form a ``channeling dip" in the outgoing flux as function of the incident beam angle $\psi$.  As pointed out first by Lindhard~\cite{Lindhard:1965}, when no slowing-down processes are involved the ``channeling" and ``blocking" dips should be identical, when compared for the same particles, energies, crystals and crystal directions.

 Channeled ions loose their energy to electrons. They penetrate distances much larger than the characteristic separation of atoms along the channels, thus they interact with hundreds or thousands of lattice atoms.  For energies in the keV range and above, channeled ions penetrate distances of at least several 10's of nm (see Appendix A, where we use the Lindhard-Scharff~\cite{Lindhard-Scharff, Dearnaley:1973} model of electronic energy loss to calculate the penetration length of ions). These are distances much longer than the separation of atoms along the channels, which are  similar to the lattice constant, i.e. approximately 0.5 nm for Si and Ge (see Appendix B).

The  potential importance of the channeling effect for direct  dark matter detection was first pointed out for NaI (Tl) by Drobyshevski~\cite{Drobyshevski:2007zj} and by the DAMA collaboration~\cite{Bernabei:2007hw}. The prospect of a daily modulation of the dark matter  signal in direct detection  due to channeling was recently  raised by  Avignone, Creswick and  Nussinov~\cite{Avignone:2008cw} in  NaI.

  In this paper we compute  the channeling fraction  of recoiling ions in Si and Ge crystals as function of the recoil energy and temperature. Si and Ge crystals are used in several direct dark matter detection experiments, such as CDMS~\cite{CDMS}, CoGeNT~\cite{CoGeNT}, Edelweiss~\cite{EDELWEISS-II}, TEXONO~\cite{TEXONO}, EURECA~\cite{EURECA}, HDMS~\cite{HDMS} and IGEX~\cite{IGEX}. In a companion paper~\cite{BGG-I} we introduced the general ideas and analytic models~\cite{Lindhard:1965, Dearnaley:1973, Gemmell:1974ub, Andersen:1967, Morgan-VanVliet, VanVliet, Andersen-Feldman, Komaki:1970, Appleton-Foti:1977, Hobler} that we use to describe these phenomena in the context of dark matter detection, and applied them to NaI (Tl). For the reader familiar with  Ref.~\cite{BGG-I} we would like to clarify which are the main differences between the calculations in Ref.~\cite{BGG-I}  and in the present paper, besides the crystal structure (see Appendix B). In this paper we use  a different expression for the continuum potentials (see Eqs. 2.1 to 2.7), which leads to a different expression for the critical channeling distance for axial channels (see Eq. 2.15). We also use a different way of deriving the critical distance for planar channels (see Eqs. 2.18 to 2.23).

 \section{Model of Channeling}

 \subsection{Continuum models}

There are different  approaches to calculate the deflections of ions traveling in a crystal.
 In  ``binary collision models''  the ion path  is computed by a computer program
(see Ref. \cite{Barrett:1971} for one of the first ones) in terms of a succession of  individual interactions, each with one of the atoms in the crystal. Crystal imperfections and lattice vibrations are thus easily and correctly taken into account. In ``continuum models'', reasonable approximations are made which allow to replace the discrete series of binary collisions with atoms by a continuous interaction between a projectile and uniformly charged strings or planes. These models allow to replace the numerical calculations by an analytic description of channeling, and provide good quantitative predictions of the behavior of projectiles in the crystal in terms of  simple physical quantities.
This is the approach we use here.

The analytical description of channeling phenomena was initially developed mostly by J. Lindhard~\cite{Lindhard:1965} and collaborators  for  ions of energy MeV and higher,  and its use was later extended to lower energies, i.e. hundreds of eV and above, mostly to apply it to ion implantation in Si.     For the low energy range, we found most useful the work of G. Hobler~\cite{Hobler}, who in 1995 and 1996 perfected and checked experimentally previous continuum model predictions~\cite{cho} for axial and planar channeling at energies in the keV to a few 100 keV range, developed  to avoid channeling in the implantation of   B, P and As atoms  in Si crystals~\cite{implantation}.  This approach must be complemented by determination of parameters through data fitting or simulations. Moreover, lattice vibrations are more difficult to include in continuum models. Since we use a continuum model, our results should in last instance be checked by using some of the many sophisticated simulation programs that  implement the binary collision approach or mixed approaches (e.g.~\cite{Monte-Carlo-programs}).

Our calculation  is based on  the classical analytic models developed in the 1960's and 70's, in particular by Lindhard~\cite{Lindhard:1965, Dearnaley:1973, Andersen:1967, Morgan-VanVliet, VanVliet, Andersen-Feldman, Komaki:1970,  Appleton-Foti:1977, Hobler}. The fact that  the de Broglie wavelengths of ions in the keV  energy  range are of the  order of $\sim$ 0.01 pm (and smaller at higher energies),  thus much shorter than the  lattice constant of a crystal ($\sim$ 500 pm, see Appendix B), justifies using a classical treatment. We use the continuum string and plane model, in which the screened Thomas-Fermi potential is averaged over a direction parallel to a row or a plane. This averaged potential  is considered to be uniformly smeared out along the row or plane of atoms, which is a good approximation if the propagating ion interacts with many lattice atoms in the row or plane by a correlated series of many consecutive glancing collisions with lattice atoms. We are going to consider just one row, which simplifies the calculations and is
correct except at the lowest energies we consider, as we explain below.

There are several good analytic approximations of the screened potential. Except when said otherwise, in this paper  we use  Moli\`{e}re's approximation, following the work of Hobler~\cite{Hobler} and Morgan and Van Vliet~\cite{Morgan-VanVliet, VanVliet}.  Moli\`{e}re's approximations  of the continuum potentials are more complicated  and also somewhat better than Lindhard's expressions,
which we used in our paper devoted to NaI~\cite{BGG-I}. Lindhard's expressions are easier to manipulate algebraically to obtain different quantities of interest. Still in this paper we used some expressions derived from
Lindhard's form of the potentials.

In  Moli\`{e}re's approximation~\cite{Gemmell:1974ub}  the axial continuum potential, as a function of the transverse distance $r$ to the string, is
 \begin{equation}
 U(r)=\left(2Z_1 Z_2 e^2/d\right)f(r/a)
 =E\psi_1^2f(r/a),
 \end{equation}
 where $E$ is the energy of the propagating particle and  $\psi_1$  is a dimensionless parameter defined by
 \begin{equation}
\psi_{1}^2=\frac{2Z_{1}Z_{2}e^2}{E d},
\label{psi1}
\end{equation}
$Z_1$, $Z_2$ are the atomic numbers of the recoiling and lattice nuclei respectively, $d$ is the spacing between atoms in the row, $a$ is the Thomas-Fermi screening distance, $a= 0.4685 {\text {\AA} } (Z_1^{1/2} + Z_2^{1/2})^{-2/3} $~\cite{Barrett:1971, Gemmell:1974ub} (1.225$\times 10^{-2}$ nm and 0.9296$\times 10^{-2}$ nm  for a Si ion in Si and a Ge ion in Ge respectively, see Appendix B) and   $E= Mv^2/2$ is the kinetic energy of the propagating ion.
 Moli\`{e}re's screening function~\cite{Gemmell:1974ub} for the continuum potential is
 \begin{equation}
 f(\xi)=\sum_{i=1}^{3}{\alpha_i K_0(\beta_i \xi)}.
 \end{equation}
Here $K_0$ is the zero-order modified Bessel function of the second kind, and the dimensionless  coefficients  $\alpha_i$ and  $\beta_i$ are  $\alpha_i=\{ 0.1, 0.55, 0.35 \}$  and $\beta_i=\{ 6.0, 1.2, 0.3 \}$ ~\cite{Watson:1958}, for $i=1,2,3$. The string of crystal atoms is at $r=0$.

 In our case, $E$ is the recoil energy imparted to the ion in a collision with a WIMP,
\begin{equation}
E = \frac{|\vec{\bf q}|^2}{2M},
\end{equation}
and $\vec{\bf q}$ is the recoil momentum.

 The continuum planar potential  in Moli\`{e}re's approximation~\cite{Gemmell:1974ub}, as a function of the distance $x$ perpendicular to the plane, is
 \begin{equation}
 U_p(x)=\left(2\pi n Z_1 Z_2 e^2 a\right)f_p(x/a)
 =E\psi_a^2f_p(x/a),
 \end{equation}
where the dimensionless parameter $\psi_a$  is defined  as
\begin{equation}
\psi_a^2=\frac{2\pi n Z_1 Z_2 e^2 a}{E},
\label{psi_a}
\end{equation}
and $n= N d_{pch}$ is the average number of atoms per unit area, where $N$ is the atomic density and $d_{pch}$ is the width of the planar channel, i.e.  the  interplanar spacing (thus, the average distance of atoms within a plane is $d_p=1/ \sqrt{N d_{pch}}$).
 The subscript p denotes ``planar" and
 \begin{equation}
 f_p(\xi)=\sum_{i=1}^{3}{(\alpha_i/\beta_i) \exp(-\beta_i \xi)},
 \end{equation}
 where the coefficients $\alpha_i$ and $\beta_i $ are the same as above.
 The plane is at $x=0$.

 Examples of axial and planar continuum potentials  for a  Si ion propagating in a Si crystal and a Ge ion propagating in a Ge crystal  are shown in Fig.~\ref{U}.

\FIGURE[h]{\epsfig{file=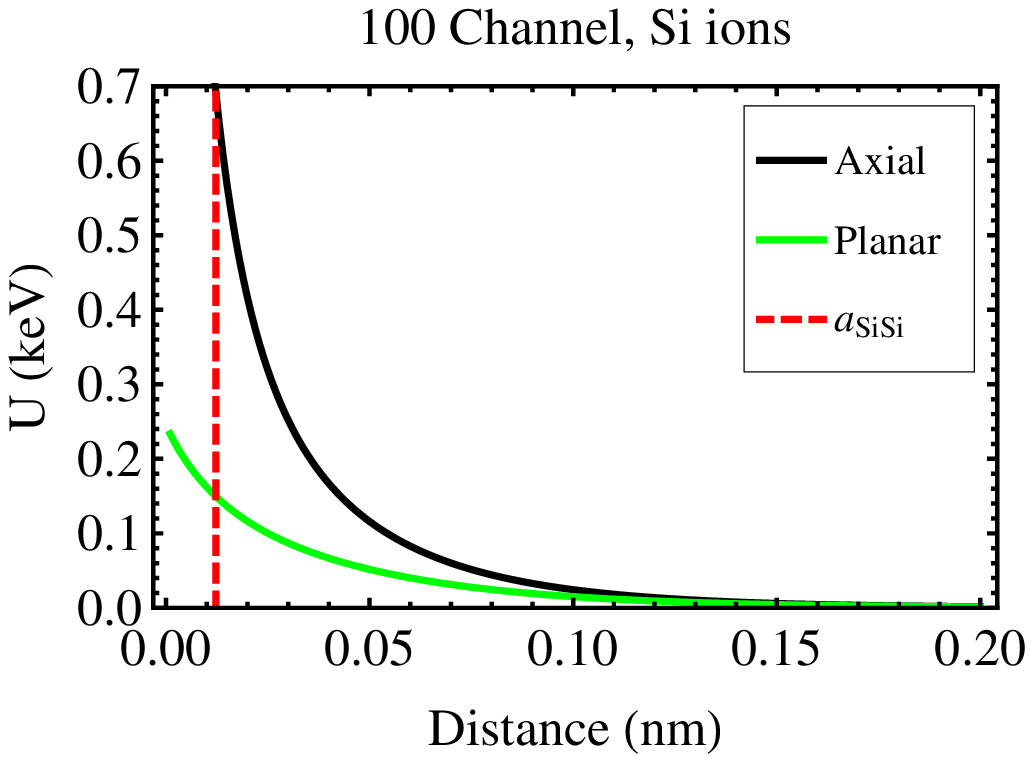,height=135pt}
\epsfig{file=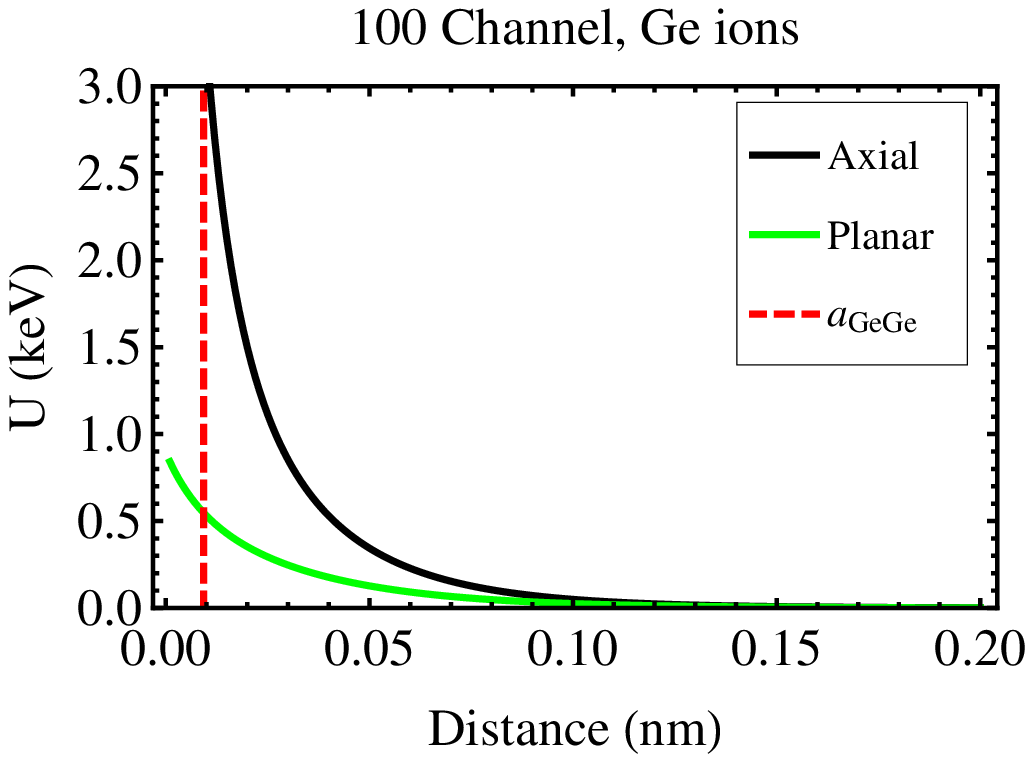,height=135pt}
        \caption{Continuum axial (black) and planar (green/gray) potentials for (a) Si and (b) Ge ions, propagating  in the $<$100$>$ axial and \{100\} planar channels of a Si or Ge crystal respectively. The screening radii shown as vertical lines are $a_{SiSi}=0.01225$ nm and $a_{GeGe}=0.009296$ nm (see Appendix B).}%
	\label{U}}

 The continuum model  does not imply that the potential energy of an ion moving near an atomic row is well  approximated by the continuum potential $U$. The actual potential consists of sharp peaks  near the atoms and deep valleys in between. The continuum model says that the net deflection due to the succession of impulses from the peaks is identical to the  deflection due to a force $-U'$. This is only so if the ion never approaches so closely  any individual atom that it suffers a large-angle collision.  Lindhard  proved that for a string of atoms this is so only if
 \begin{equation}
U''(r) <  \frac{8}{d^2} E ,
\label{U''}
\end{equation}
where the double prime denotes the second derivative with respect to $r$. Replacing the inequality in Eq.~\ref{U''} by an equality defines an energy dependent critical distance $r_c$ such that  $r >  r_c$ for the continuum model to be valid. Morgan and Van Vliet~\cite{Morgan-VanVliet} also  derived a condition for axial channels, similar to Eq.~\ref{U''} but with the factor 8 replaced by 16.

The breakdown of the continuum theory for a planar channel is more involved than for an axial channel because  the atoms in the plane contributing to the scattering of the propagating ion are usually displaced laterally within the plane with respect to the ion's trajectory. Thus the moving ion does not encounter atoms at a fixed separation or at fixed impact parameter as is the case for a row.  Morgan and Van Vliet~\cite{Morgan-VanVliet} reduced the problem of scattering from a plane of atoms to the scattering of an equivalent row of atoms contained in a strip centered on the projection of the ion path on the plane of atoms. They then  applied Eq.~\ref{U''} as the condition for planar channeling   to the fictitious string defined in this way (more about this below).
\subsection{The transverse energy}

Lindhard proved that for channeled particles the longitudinal component   $v \cos\phi$ of the velocity, i.e. the component along the direction of the row or plane of the velocity, may be treated as constant (if energy loss processes are neglected). Then, in the continuum model, the trajectory of the ions can be completely described in terms of the transverse direction, perpendicular to the row or plane considered. For small  angle $\phi$  between the ion's trajectory and  the atomic  row (or plane)
in the direction perpendicular to  the row (or plane), the  so called ``transverse energy"
\begin{equation}
E_{\perp} = E \sin^2\phi + U\simeq E \phi^2 + U
\label{E-perp}
\end{equation}
is conserved. In Eq. \ref{E-perp} relativistic corrections are neglected.

 Let  $r_i$ be the initial position at which the WIMP nucleus collision occurs, i.e. if $r_i >0$ the recoiling nucleus was displaced with respect to its position of equilibrium in a crystal row when it collided with a WIMP. We call $\phi_i$ the angle of the initial recoil momentum  with respect to the row of atoms and $E$  the initial recoil energy  of the propagating ion.  Given  these initial parameters, the issue of  where to define $E_{\perp}$ arises. Namely, we define
\begin{equation}
E_{\perp}= E \sin\phi_i^2 + U(r^*),
\label{E-perp-HP}
\end{equation}
but there are different possible choices for $r^*$, the position at which to measure the potential $U$.
 In our case, the recoiling ion leaves an empty lattice site, thus it moves away from an empty lattice site in the potential generated by its neighboring lattice atoms.
 So the potential the recoiling ion moves through at the moment of collision is very small, and
the recoiling ion conserves its momentum and direction of motion until it gets very near the nearest neighbor, a distance $d$ away along the string. At this moment, it is at a distance
\begin{equation}
r^* \equiv  r_i + d \tan{\phi_i}
\label{r*definition}
\end{equation}
from its nearest neighbor.
Therefore, as we did in Ref.~\cite{BGG-I}, we will  make the approximation of defining the potential entering into Eq.~\ref{E-perp-HP} at  this  position $r^*$.

\subsection{Minimum distances of approach and critical channeling angles}

The conservation of the transverse energy provides a definition of the minimum distance of approach to the string, $r_{\rm min}$ (or to the plane of atoms $x_{\rm min}$), at which the trajectory of the ion makes  a zero angle with the string (or plane), and also of the angle $\psi$ at which the ion exits from the string (or plane), i.e. far away from it where $U \simeq0$. In reality the furthest position from   a string or plane of atoms is  the middle of the channel, whose width we call  $d_{ach}$ for an axial channel ($d_{pch}$ for a planar channel).  Thus, for an axial channel
\begin{equation}
\label{eq:consetrans}
E_{\perp}= U(r_{\rm min}) =  E \psi^2 +U(d_{ach}/2).
\end{equation}
We proceeded in two ways to define the axial channel radius $(d_{ach}/2)$ for the axial channels we included in our calculation. We used the contour plots of the axial continuum potentials plotted in a plane perpendicular to the channels shown in Fig.~3 of the 1995 paper of Hobler~\cite{Hobler} to read off the  channel radius $d_{\rm ach}/2$ of the $<$100$>$, $<$110$>$ and  $<$111$>$ axial channels in terms of the lattice constant   a$_{lat}$. They are  0.25 a$_{lat}$, 0.375 a$_{lat}$, and $ \sqrt{0.2^2+0.12^2}$ a$_{lat}=$ 0.233 a$_{lat}$, respectively. For the other axial channels we considered, $<$211$>$ and $<$311$>$, we define the channel width $d_{\rm ach}$ in terms of the interatomic distance $d$ in the corresponding row as $d_{ach}= 1/ \sqrt{N d}$, where is $N$ the atomic density. For a planar channel we replace the axial potential at the middle of the axial channel  $U(d_{ach}/2)$ in Eq.~\ref{eq:consetrans}  by  the planar potential at the middle of the planar channel  $U_p(d_{pch}/2)$ (the channel width $d_{pch}$ was defined after Eq.~\ref{psi_a}).

For axial channeling Lindhard equates the  condition for channeling with  the condition in  Eq.~\ref{U''} for the validity of the continuum model.
Replacing the inequality in Eq.~\ref{U''} by an equality defines an energy dependent critical distance $r_c$, so that channeling can happen only if the propagating ion always keeps a distance $r > r_c$.
 Morgan and Van Vliet~\cite{Morgan-VanVliet} use 5 instead of 8 in Eq.~\ref{U''}, because this agrees better with their simulations of channeling in copper crystals. Following Hobler~\cite{Hobler}, we use here Morgan and Van Vliet's equation to define $r_c$, i.e.
 \begin{equation}
U''(r_c) =  \frac{5}{d^2} E.
\label{MVU''}
\end{equation}
 With Moli\`{e}re's form of the potential it is not possible to solve analytically for $r_c$.
 Morgan and Van Vliet~\cite{Morgan-VanVliet} gave the following approximate analytical solution for the axial channeling minimum distance of approach,
\begin{equation}
r^{MV}_c= (2/3) a \sqrt{\alpha} \left[1-(\sqrt{\alpha}/19) + (\alpha/700)\right]
\label{rcritMoliere}
\end{equation}
\FIGURE{\epsfig{file=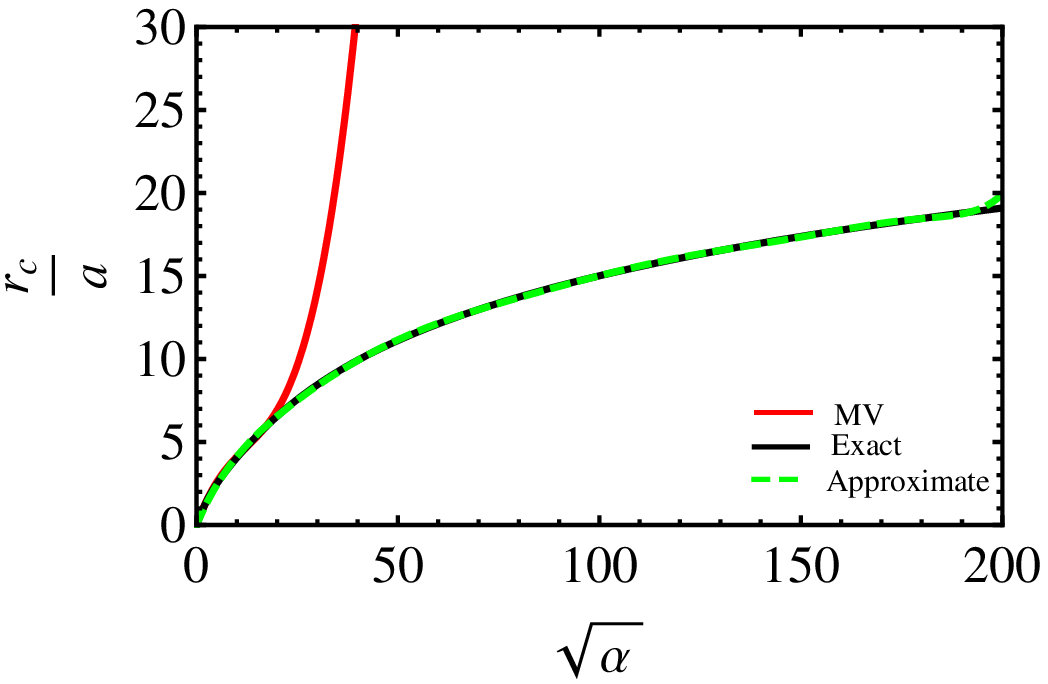,height=140pt}
\epsfig{file=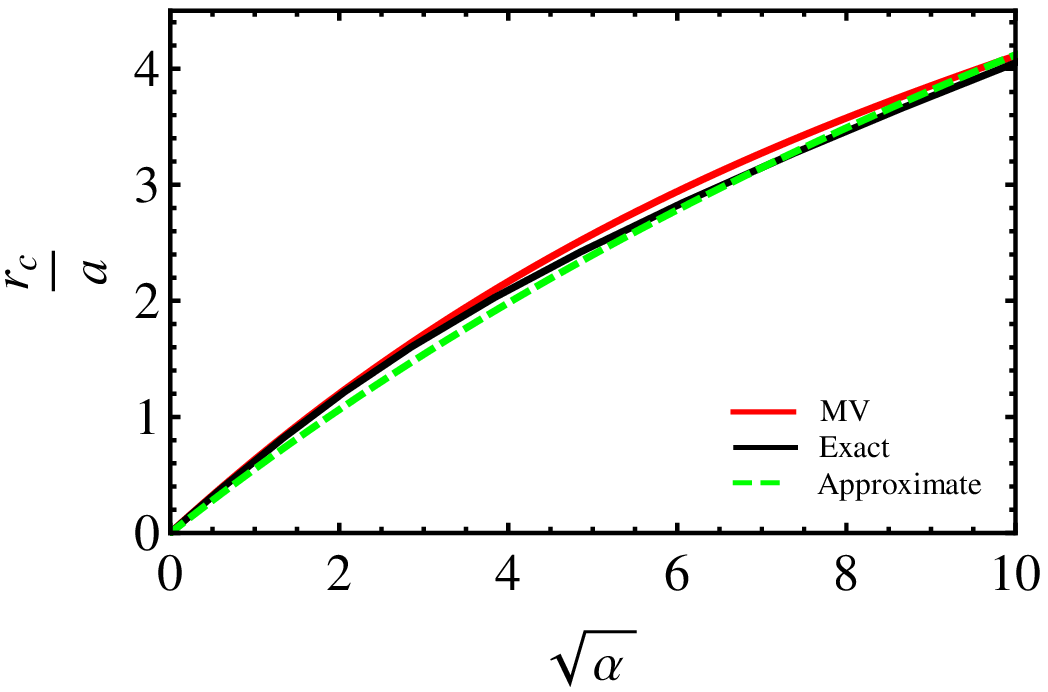,height=135pt}
        \caption{Comparison of the exact numerical solution (solid black) of Eq.~\ref{MVU''} for the
  critical distance of approach $r_c(E)$ and the approximate analytic expression in Eq.~\ref{rcritPoly} (dashed green) as a function of $\sqrt{\alpha}= \sqrt{Z_1 Z_2 e^2 d/ a^2 E}$ for  (a) the high $\sqrt{\alpha}$ (low energy) range, and (b)  the low $\sqrt{\alpha}$ (high energy) range. The Morgan and Van Vliet approximation to $r_c(E)$ in Eq.~\ref{rcritMoliere} is also shown (solid red- labeled MV).}
\label{rcPoly}}
with $\alpha= (Z_1 Z_2 e^2 d/ a^2 E)$.  This solution is not correct at low energies (high values of $\alpha$).  As can be seen in Fig.~\ref{rcPoly} (and also in Figs.~8 and 13 of the 1995 paper of Hobler~\cite{Hobler})  the steep increase in the approximate Morgan and Van Vliet solution at low energies (see the curve labeled ``MV" in Fig.~\ref{rcPoly}.a) is not present in  the numerical solution (see the curve labeled ``Exact" in Fig.~\ref{rcPoly}.a) of  $r_c$. Instead of Eq.~\ref{rcritMoliere} we use here a better approximate analytic solution  obtained by fitting a degree nine polynomial to the exact solution of Eq.~\ref{MVU''},
\begin{eqnarray}
r_c &=& a~[0.57305 \sqrt{\alpha}  - 0.0220301 (\sqrt{\alpha}) ^2 + 0.000728889 (\sqrt{\alpha}) ^3\nonumber\\
&&- 0.0000155189 (\sqrt{\alpha})^4 + 2.04162 \times 10^{-7} (\sqrt{\alpha})^5 - 1.65057 \times 10^{-9} (\sqrt{\alpha})^6\nonumber\\
&&+ 7.9749 \times 10^{-12} (\sqrt{\alpha})^7 -2.11041 \times 10^{-14} (\sqrt{\alpha})^8 + 2.35121 \times 10^{-17} (\sqrt{\alpha})^9].
\label{rcritPoly}
\end{eqnarray}
 Eq.~\ref{rcritPoly} is valid from $E$  of 1 keV to 29 TeV ( which corresponds to values of $\sqrt\alpha$ between 180 and 0.000158). Fig.~\ref{rcPoly}  shows  a comparison of the exact numerical solution  $r_c(E)$ of Eq.~\ref{U''} and the approximate analytic solution Eq.~\ref{rcritPoly} as a function of $\sqrt{\alpha}$ (divided by the screening distance $a$). The high  and low $\sqrt{\alpha}$ range in Fig.~\ref{rcPoly}.a and \ref{rcPoly}.b respectively corresponds to low  and high energies. The maximum percentage error between the exact solution and the  analytic approximation we use is 11.5 \%.

 Fig.~\ref{rcPoly-Si-Ge} shows the critical distance of approach $r_c(E)$  in Eq.~\ref{rcritPoly} as a function of energy of the propagating ion for several axial channels, for  Si ions propagating in a Si crystal and Ge ions propagating in a Ge crystal.

Since $r_c$ is the smallest possible minimum distance of approach to the string of a channeled propagating ion  for a given energy $E$, i.e. $r_{\rm min} > r_c$, and  the potential $U(r)$  decreases monotonically with increasing $r$, then
\begin{equation}
U(r_{\rm min}) < U(r_c).
\label{defrcrit}
\end{equation}
Using Eq.~\ref{eq:consetrans}, this can be further translated into an upper bound on $E_{\perp}$ and thus on  $\psi$, the angle the ion makes with the string far away from it,
\begin{equation}
\psi < \psi_{c}(E)=  \sqrt{ \frac{U(r_c(E))- U(d_{ach}/2)}{E} }.
\label{psicritaxial}
\end{equation}
 $\psi_{c}(E)$ is the critical channeling  angle for the particular axial channel, i.e. it  is the maximum angle  the  propagating ion can make with the string far  away from it (in the middle of the channel) if the ion is channeled.
 \FIGURE{\epsfig{file=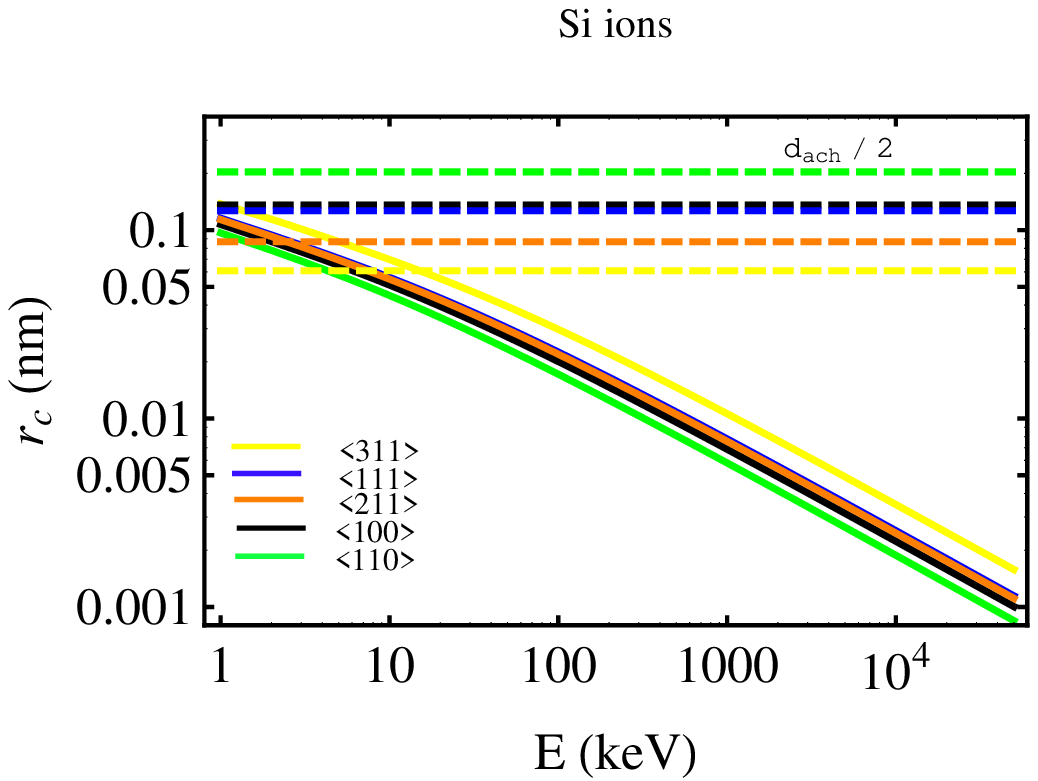,height=160pt}
 \epsfig{file=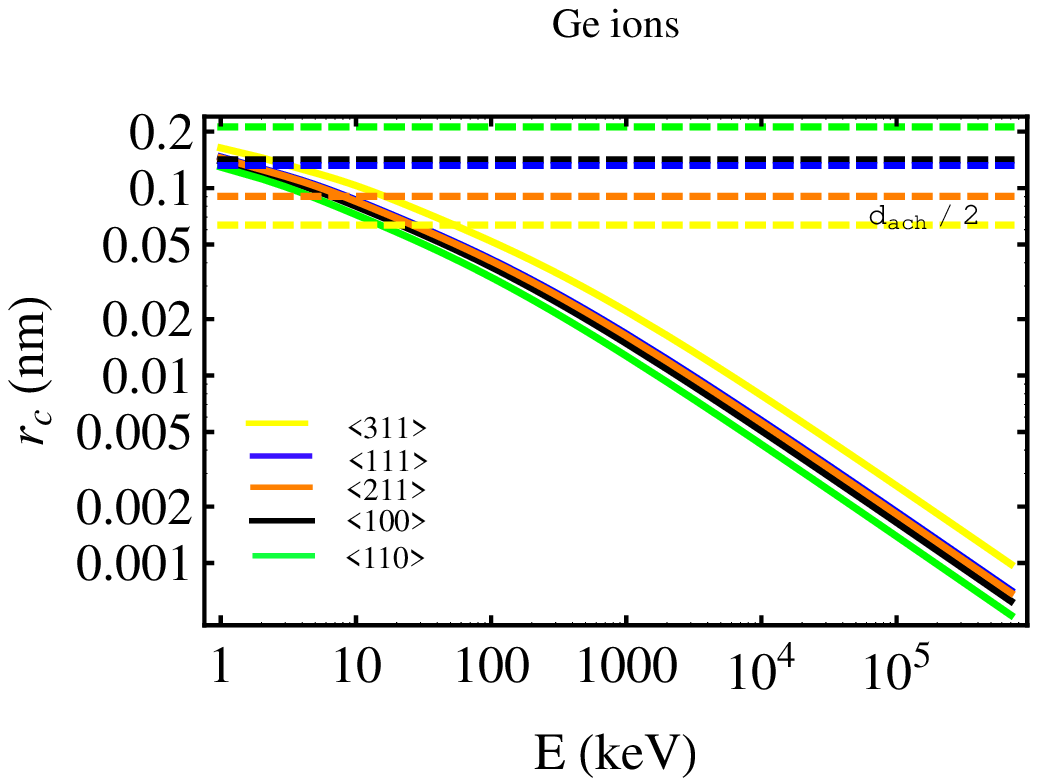,height=160pt}
        \caption{Critical channeling distance of approach $r_c(E)$  in Eq.~\ref{rcritMoliere}  as a function of energy of the propagating ion for several axial channels, for (a)  Si ions propagating in a Si crystal and (b) Ge ions propagating in a Ge crystal.}%
	\label{rcPoly-Si-Ge}}

The critical distance  $r_c(E)$ increases as $E$  decreases (see Figs.~\ref{rcPoly-Si-Ge}, \ref{xcPoly-Si-Ge} and \ref{rc-psic-MV-100-Si} to \ref{rc-psic-MV-100-Si-c2}). At low enough $E$,  $r_c(E)$ becomes close to the radius of the channel $d_{\rm ach}/2$, and  the critical angle $\psi_{c}(E)$ (which is the maximum angle for channeling in the middle of the channel) goes to zero (see Figs.~\ref{rc-psic-MV-100-Si} to \ref{rc-psic-MV-110-Ge} and \ref{T-depent-psic-Si}, \ref{T-depent-psic-Ge}). This means that there is a minimum energy below which channeling cannot happen, even for ions moving initially in the middle of the channel. This is a reflection of the fact that the range of
the interaction between ion and lattice atoms increases with decreasing energy and at some point there is no position in the crystal where the ion would not be deflected at large angles. The existence of a minimum energy for channeling was found by  Rozhkov and DyulÕdya~\cite{Rozhkov} in 1984 and later by Hobler~\cite{Hobler} in 1996. It is clear that to compute  $r_c(E)$ when it is not small with respect to the radius of the channel $d_{\rm ach}/2$, and thus to compute the actual minimum energy for channeling, we would need to consider the effect of more than one row or plane (as done in Refs.~\cite{Rozhkov} and \cite{Hobler}), thus our results are approximate in this case.

For planar channeling we will follow the procedure of defining a  ``fictitious row" introduced by Morgan and Van Vliet~\cite{Morgan-VanVliet, Hobler}.
They reduced the problem of scattering from a plane of atoms to the scattering of an equivalent row of atoms contained in a strip of width 2$R$ ($R$ is defined below) centered on the projection of the ion path onto the plane of atoms, and took the average area per atom  in the plane, $1/N d_{pch}$ to be  2$R$ times the characteristic distance $\bar{d}$ between
atoms along this fictitious row, i.e.
\begin{equation}
\bar{d} = 1/ (N d_{pch} 2 R).
\label{eq: dbar}
\end{equation}
 Once the width $2R$ of the fictitious row is specified, one uses the channeling condition for the continuum string model, Eq.~\ref{U''}, with the average atomic composition of the plane. For $R$, Morgan and Van Vliet used the impact parameter in an ion-atom collision corresponding to a deflection angle of the order of ``the break-through" angle $\sqrt{U_p(0)/E}$. This is the minimum angle at which an ion of energy $E$ must approach the plane from far away (so that the initial potential can be neglected) to overcome the potential barrier at the center of the plane at $x=0$ (namely, so that $E_\perp= U_p(0)$).  For small scattering angles, the deflection angle $\delta$ is related to the impact parameter, in this case $R$, as (see e.g. Eq. $2.1'$ of Lindhard~\cite{Lindhard:1965})
\begin{equation}
2 E \delta= -d~U'(R),
\label{U'}
\end{equation}
where $U'$ is the derivative of the axial continuum potential, and Morgan and Van Vliet define $R$ by taking $\delta=\sqrt{U_p(0)/E}$.
Using the Moli\`{e}re's approximation for the potentials,
 Morgan and Van Vliet found the following expressions for $R$
\begin{equation}
R^{MV}= a \left(\frac{A}{2}\right) \ln\left(B~Z_1 Z_2 e^2/ a \sqrt{E U_p(0)}\right)
\label{MVR}
\end{equation}
which lead to the  $\bar{d}$ value
\begin{equation}
\bar{d}^{MV}= \left[A~a N d_{\rm pch} \ln\left(B~Z_1 Z_2 e^2/ a \sqrt{E U_p(0)}\right)\right]^{-1},
\label{dbar-MV}
\end{equation}
with coefficients $A=1.2$ and $B=4$. Morgan and Van Vliet~\cite{Morgan-VanVliet} found discrepancies with this theoretical formula  in simulations of binary collisions  of 20 keV protons in a copper crystal and adjusted the coefficients to  $A= 3.6$ and $B=2.5$. Hobler~\cite{Hobler} used both sets of coefficients and compared them with simulations and data of B and P ions propagating in Si for energies  of about 1 keV and above. Hobler concluded that the original theoretical formula was better in his case (although Hobler proposed yet another empirical relation to define $\bar{d}$).  While Eq.~\ref{U'}  seems to provide a good condition for $R$, there is a channel dependent   energy upper limit of applicability of  its approximate analytical solution in Eq.~\ref{dbar-MV},  because the logarithm in $\bar{d}^{MV}$ approaches zero as $E$ approaches  $(4 Z_1 Z_2 e^2/ a)^2/U_p(0)$. Close to this value of  $E$ there is an unphysical  fast increase in  $\bar{d}^{MV}$ (and consequently in $x_c(E)$) that indicates the break-down of the approximate solution  $\bar{d}^{MV}$ in  Eq.~\ref{dbar-MV} (and, as shown in  Fig.~13 of the 1995 paper of Hobler~\cite{Hobler}, is not found in other expressions of $x_c$).

We decided to keep  the Morgan and Van Vliet definition for $R$ in Eq.~\ref{U'}  and use the following approximate analytical solution obtained by fitting a degree five polynomial in $\ln{y}$ to the exact numerical solution of Eq.~\ref{U'}
\begin{eqnarray}
R&=& a~\big(0.716014 + 0.510922 \ln{y} + 0.12047 (\ln{y})^2 + 0.0180492 (\ln{y})^3\nonumber\\
&&+ 0.00442459 (\ln{y})^4 - 0.000824744 (\ln{y})^5 \big),
\label{R}
\end{eqnarray}
where $y=Z_1 Z_2 e^2/a \sqrt{E U_p(0)}$.

  Fig.~\ref{dbar}  shows  a comparison of the exact numerical solution of Eq.~\ref{U'}  for $R$  and its  analytical approximation in Eq.~\ref{R} (divided by $a$) as a function of $y$. Also the approximate expression of Morgan and Van Vliet in Eq.~\ref{dbar-MV} is shown in Fig.~\ref{dbar} (labeled MV).  The high  and low $y$ ranges in  Fig.~\ref{dbar}.a  and b respectively corresponds to low  and high energies. The approximate solution is not valid at  $y<0.15$ which corresponds to  $E>50$ MeV for Si, and $E>700$ MeV for Ge. Within its range of validity, the percentage error of the analytic approximation in Eq.~\ref{R} is less than 9\%.
  \FIGURE{\epsfig{file=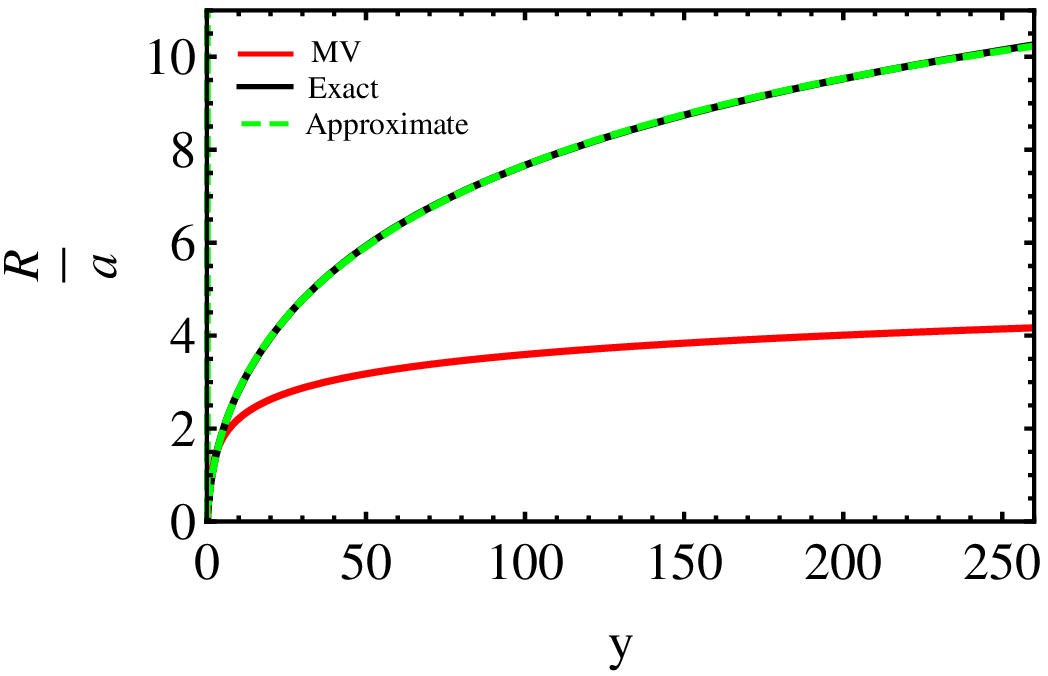,height=135pt}
 \epsfig{file=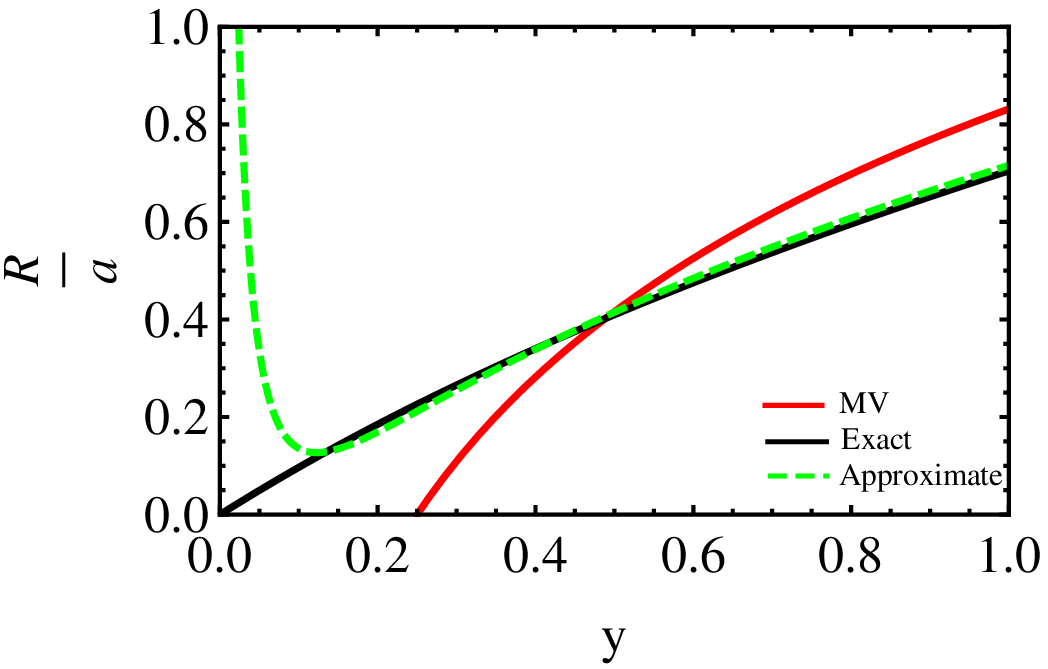,height=142pt}
        \caption{Comparison of the exact solution (solid black) of Eq.~\ref{U'} for  $R/a$   and its analytical approximation  in Eq.~\ref{R} (dashed green) as a function of  $y=Z_1 Z_2 e^2/a \sqrt{E U_p(0)}$ for the (a) high $y$ (low $E$) range and the (b) low $y$ (high $E$) range. Also the Morgan and Van Vliet  approximation to $R/a$ in Eq.~\ref{MVR} is shown (solid red- labeled MV).}%
	\label{dbar}}

   Let us call $\bar{r}_c(E)$ the critical distance obtained from  Eq.~\ref{rcritPoly} for the fictitious row, whose interatomic distance is $\bar{d}$ in Eq.~\ref{eq: dbar} in which the distance $R$ is given in Eq.~\ref{R}. Then, the minimum distance of approach for planar channeling is
\begin{equation}
x_c(E) \equiv \bar{r}_c(E).
\label{ourxcrit}
\end{equation}
Fig.~\ref{xcPoly-Si-Ge} shows the plot of  $x_c(E)$  (obtained from using Eq.~\ref{R} for the fictitious string) as a function of energy for  the most important planar channels, i.e.  \{100\}, \{110\}, \{111\}, \{210\} and \{310\}. Fig.~\ref{xcPoly-Si-Ge} shows that we can safely extend our approximation to 50 MeV for Si ions in a Si crystal and to 700 MeV for Ge ions in Ge crystals.
   \FIGURE{\epsfig{file=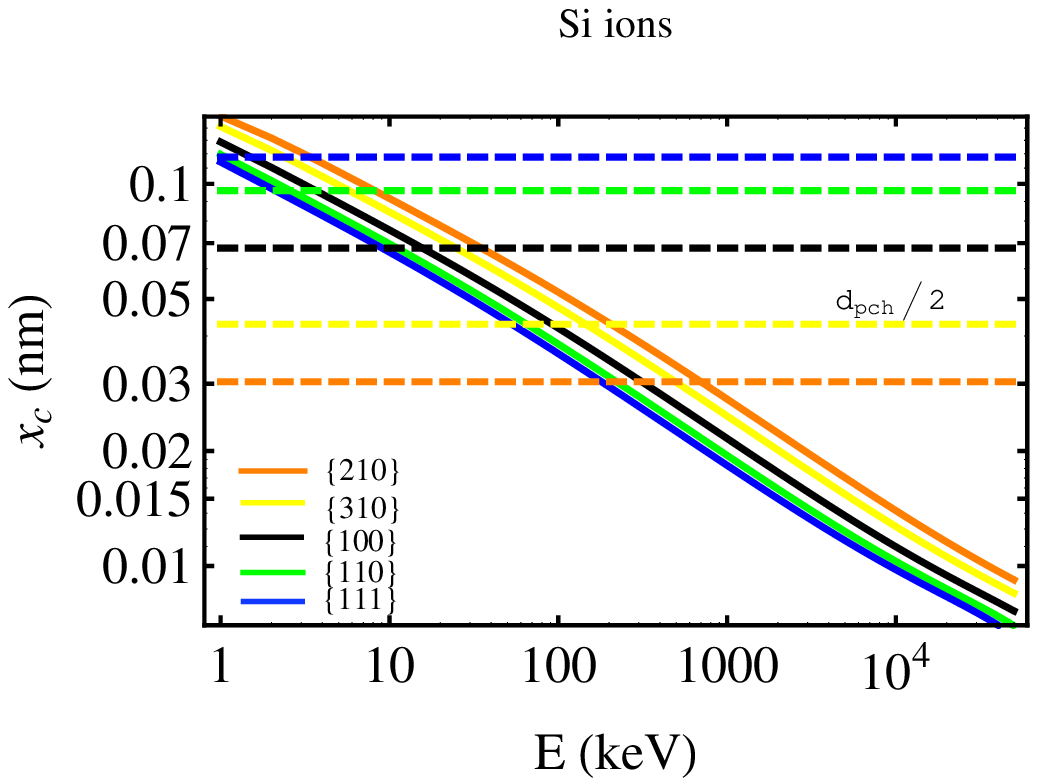,height=160pt}
 \epsfig{file=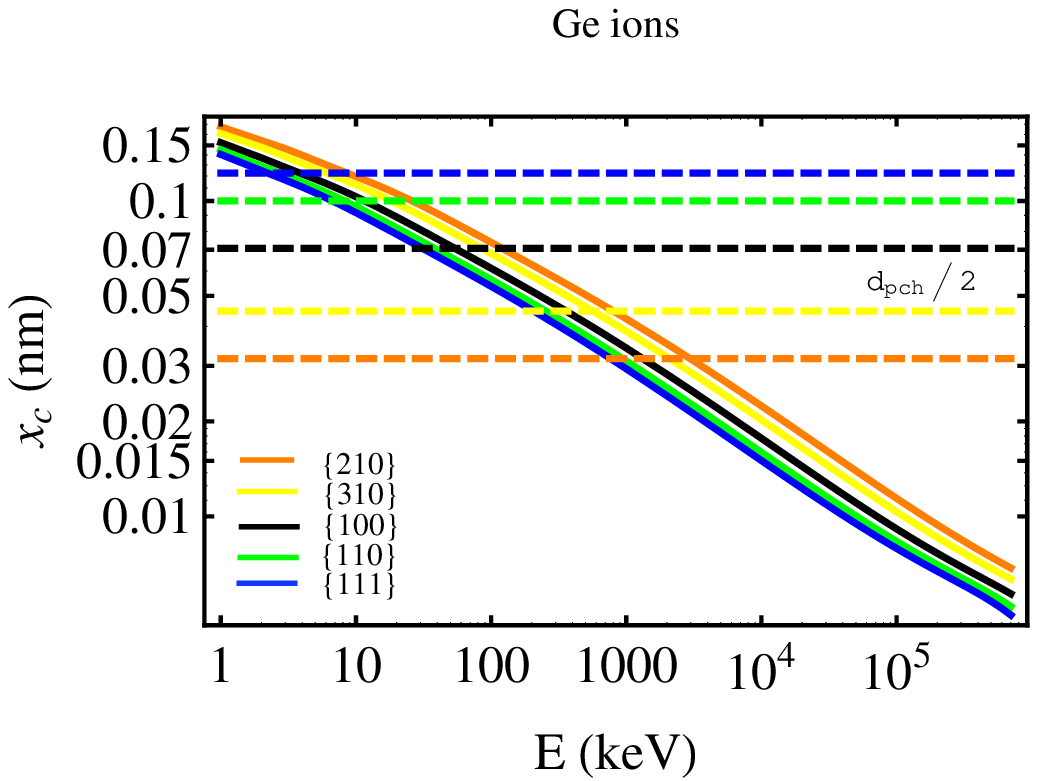,height=160pt}
        \caption{Critical channeling distances $x_c(E)$ in Eq.~\ref{ourxcrit}  as a function of the energy  of the propagating ion for different planar channels, for (a) Si ions propagating in a Si crystal and (b) Ge ions in Ge.}%
	\label{xcPoly-Si-Ge}}

Writing equations equivalent to Eqs.~\ref{eq:consetrans} and \ref{defrcrit} for planar channels, namely
\begin{equation}
E_{\perp}= U_p(x_{\rm min}) =  E (\psi^p)^2 +U_p(d_{pch}/2)
\label{planarEperp}
\end{equation}
and
\begin{equation}
U_p(x_{\rm min}) < U_p(x_c(E)),
\label{planarmincond}
\end{equation}
we obtain an equation similar to Eq.~\ref{psicritaxial} but for the maximum planar channeling angle, the critical planar channeling angle
\begin{equation}
\psi^p_c(E)=  \sqrt{ \frac{U_p(x_c(E))- U_p(d_{pch}/2)}{E} }.
\label{psicritplanar}
\end{equation}
For very small energies, for which $x_c(E) \geq d_{\rm pch}/2$ no channeling is possible (the maximum distance  to any plane cannot be larger than half the width of the channel separating them) and $\psi^p_c=0$ (see Figs.~\ref{xcPoly-Si-Ge}, \ref{rc-psic-MV-100-Si} to \ref{rc-psic-MV-110-Ge} and \ref{T-depent-psic-Si}.b, \ref{T-depent-psic-Ge}.b). When $x_c(E)$
 approaches the middle of the channel the effect of other planes should be considered, so our approximation of using the potential of only one plane is not correct in this regime.

The static lattice critical distances presented in  Figs~\ref{rcPoly-Si-Ge} and \ref{xcPoly-Si-Ge} (also in the  left panels of  Figs. \ref{rc-psic-MV-100-Si}, \ref{rc-psic-MV-110-Si}, \ref{rc-psic-MV-100-Ge} and \ref{rc-psic-MV-110-Ge})  do  not include  thermal effects. These are important and must be taken into account. They increase the critical channeling distances and consequently decrease the critical channeling angles as the temperature increases (as clearly shown in  Fig.~\ref{rc-psic-MV-100-Si-c1} and \ref{rc-psic-MV-100-Si-c2}).
     \FIGURE{\epsfig{file=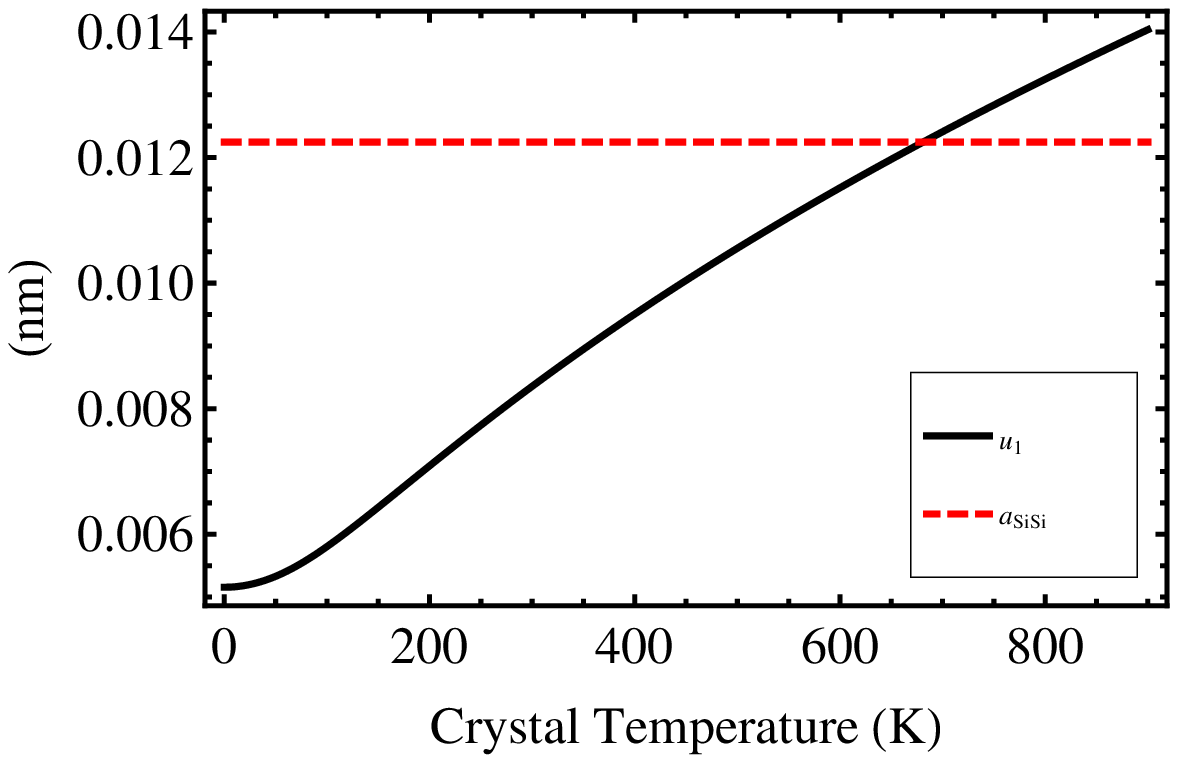,height=135pt}
 \epsfig{file=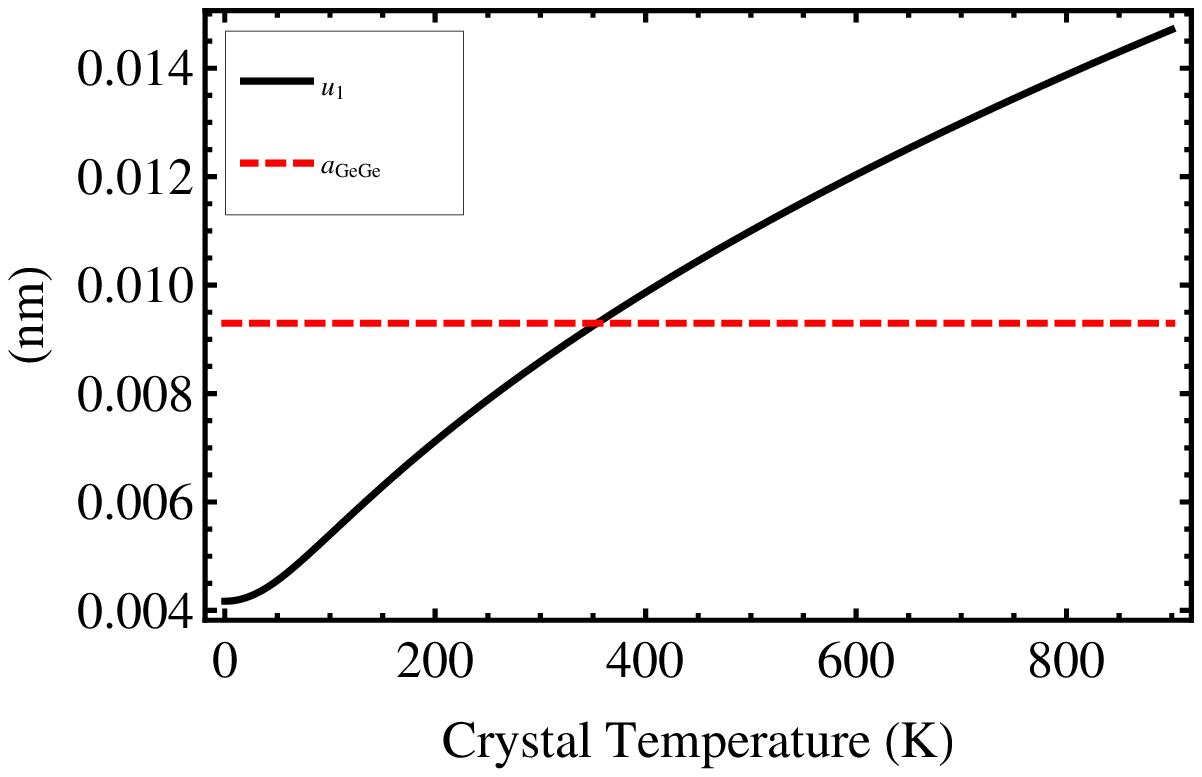,height=135pt}
        \caption{Temperature dependent Debye model  one dimensional rms vibration amplitude $u_1(T)$ (Eq.~\ref{vibu1})  of the atoms in (a)  a Si crystal and (b)  a Ge crystal. For comparison the  Thomas-Fermi screening distances for two Si atoms and  two Ge  atoms,  $a_{SiSi}$ and $a_{GeGe}$ respectively are also indicated (see App. A).}%
	\label{figu1}}

\subsection{Temperature dependent critical distances and angles}

So far we have been considering static strings and planes, but the atoms in a crystal are actually vibrating.  We use here the Debye model to take into account the zero point energy and thermal vibrations of the atoms in a crystal.  The one dimensional rms vibration amplitude $u_1$ of the atoms in a crystal in this model is~\cite{Gemmell:1974ub, Appleton-Foti:1977}
\begin{equation}
u_1(T)=12.1 \, \text{\AA} \, \left[\left(\frac{\Phi(\Theta/T)}{{\Theta/T}} + \frac{1}{4}\right)(M\Theta)^{-1}\right]^{1/2},
\label{vibu1}
\end{equation}
where the 1/4 term accounts for the zero point energy, $M$ is the atomic mass in amu, $\Theta$ and $T$ are the Debye temperature and the temperature of the crystal in K, respectively, and $\Phi(x)$ is the Debye function,
\begin{equation}
\Phi(x)=\frac{1}{x}\int_{0}^{x}{\frac{t dt}{e^t -1}}.
\end{equation}
The Debye temperatures of Ge and Si are respectively  $\Theta=290\;^\circ$K and  $\Theta=490\;^\circ$ K ~\cite{Gemmell:1974ub, Hobler}.  The vibration amplitude $u_1$ as a function of the temperature $T$ is plotted in Fig.~\ref{figu1} for Si and Ge crystals.  At room temperature (20 $^\circ$C), $u_1=0.00849$ nm  for Ge and $u_1=0.00827$ nm for  Si.
     \FIGURE{\epsfig{file=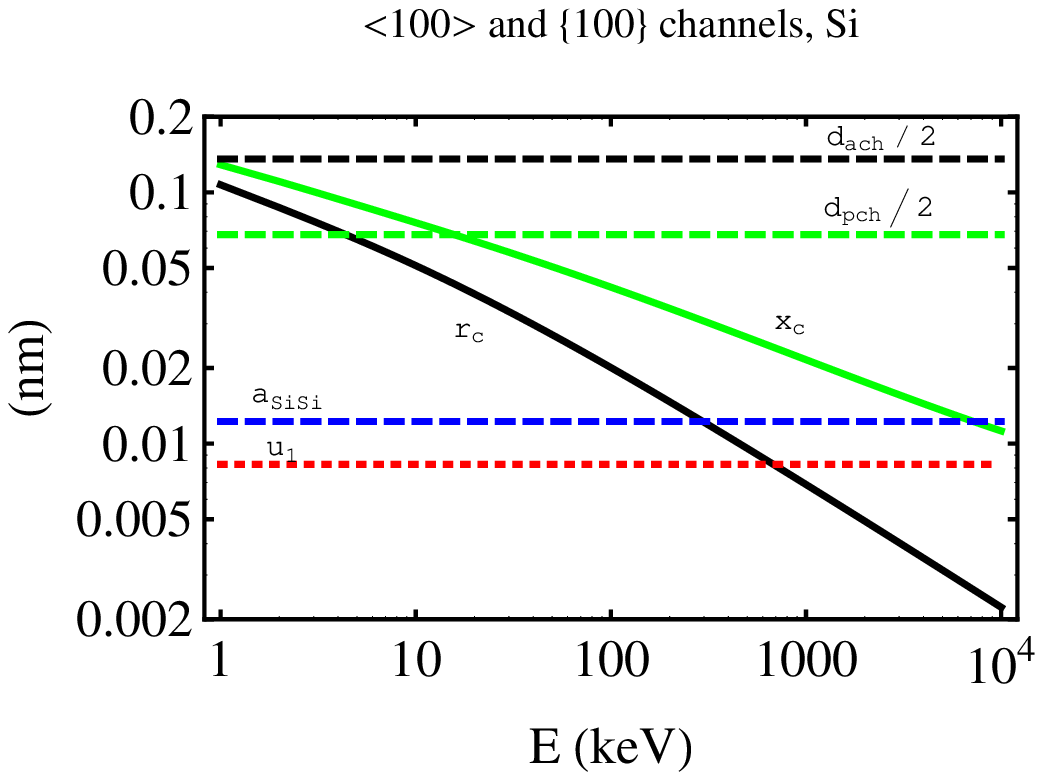,height=160pt}
 \epsfig{file=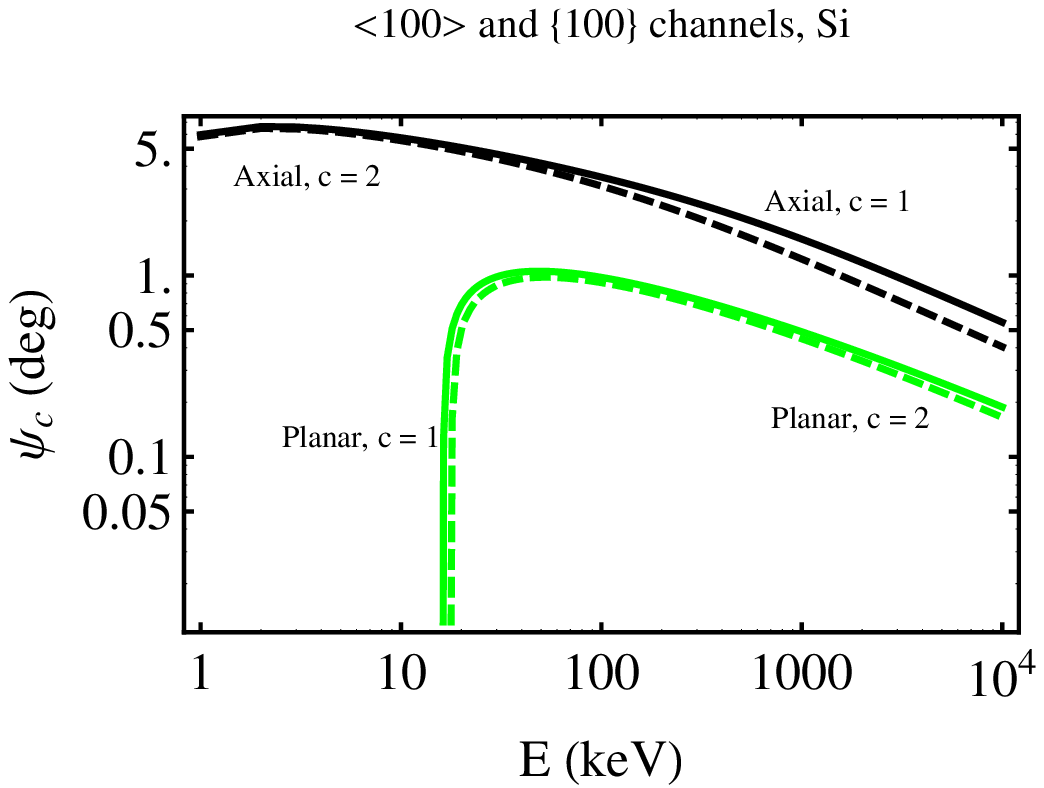,height=160pt}
        \caption{(a) Static critical distances of approach and Debye one dimensional rms vibration amplitude $u_1$ of the atoms in the crystal at 20 $^o$C and (b) critical channeling angles at 20 $^\circ$C with temperature effects computed assuming $c_1=c_2=c$ and $c=1$ or $c=2$ as indicated, as a function of the energy of propagating Si ions  in the $<$100$>$ axial (black) and \{100\} planar (green/light gray) channels
  of a  Si crystal.}%
	\label{rc-psic-MV-100-Si}}
     \FIGURE{\epsfig{file=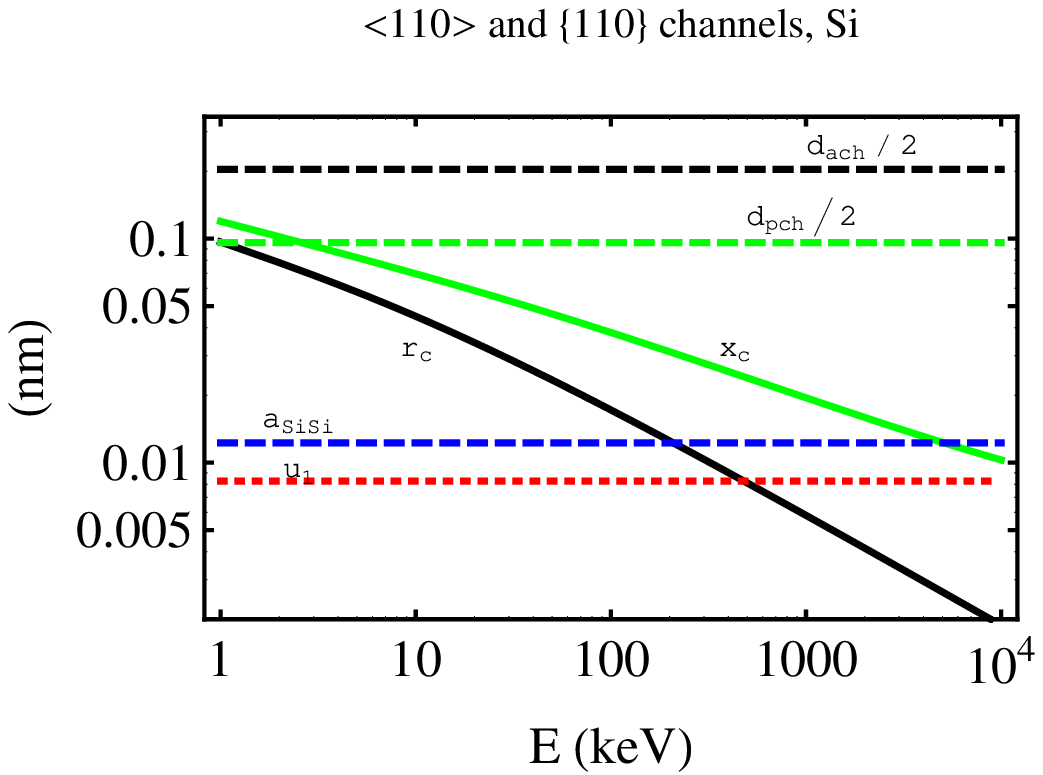,height=160pt}
 \epsfig{file=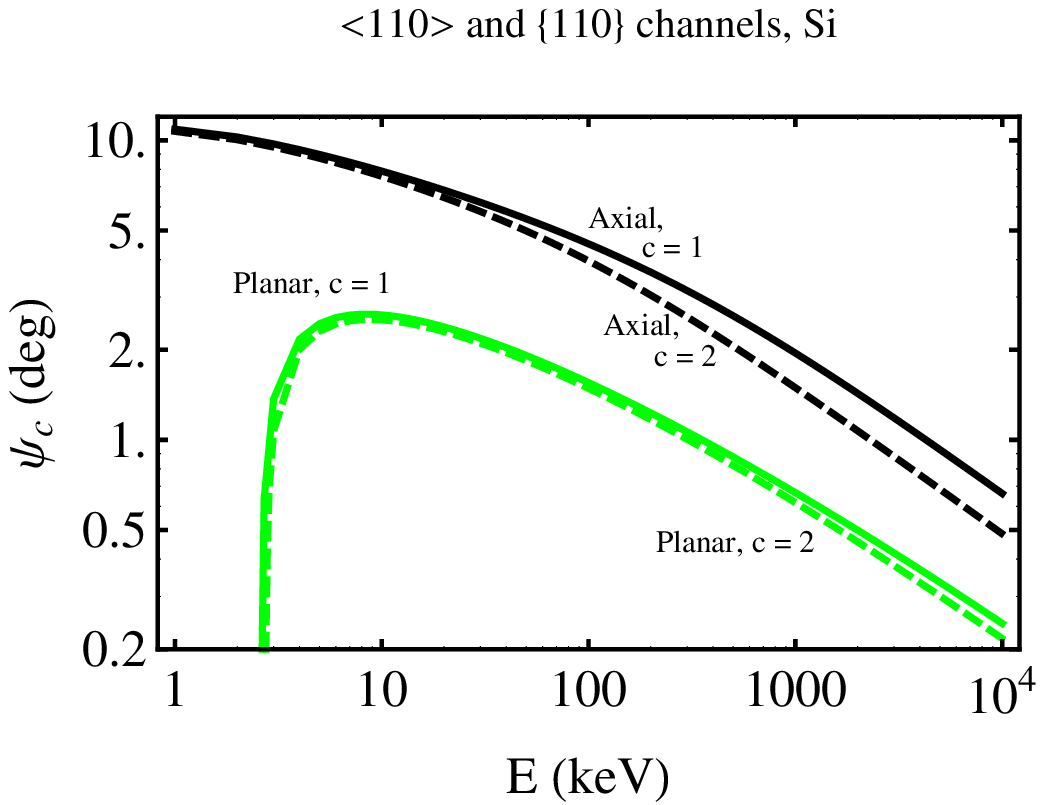,height=160pt}
        \caption{Same as in Fig.~\ref{rc-psic-MV-100-Si} but for the $<$110$>$ axial and \{110\} planar channels of a  Si crystal.}%
	\label{rc-psic-MV-110-Si}}
     \FIGURE{\epsfig{file=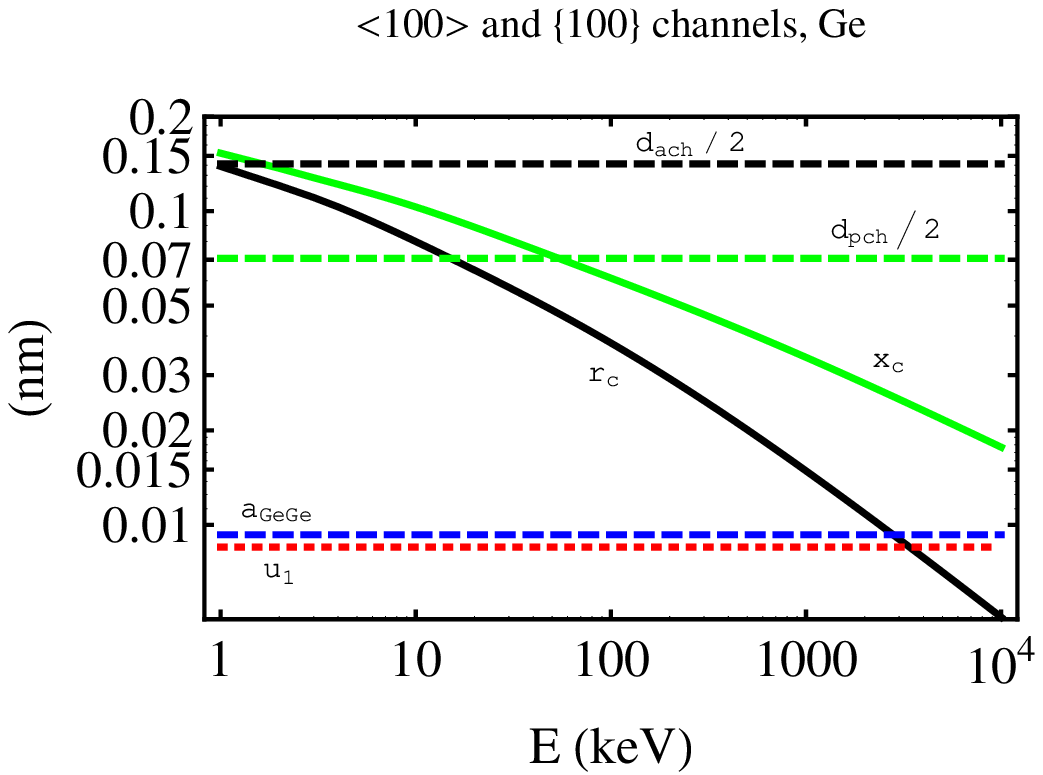,height=160pt}
 \epsfig{file=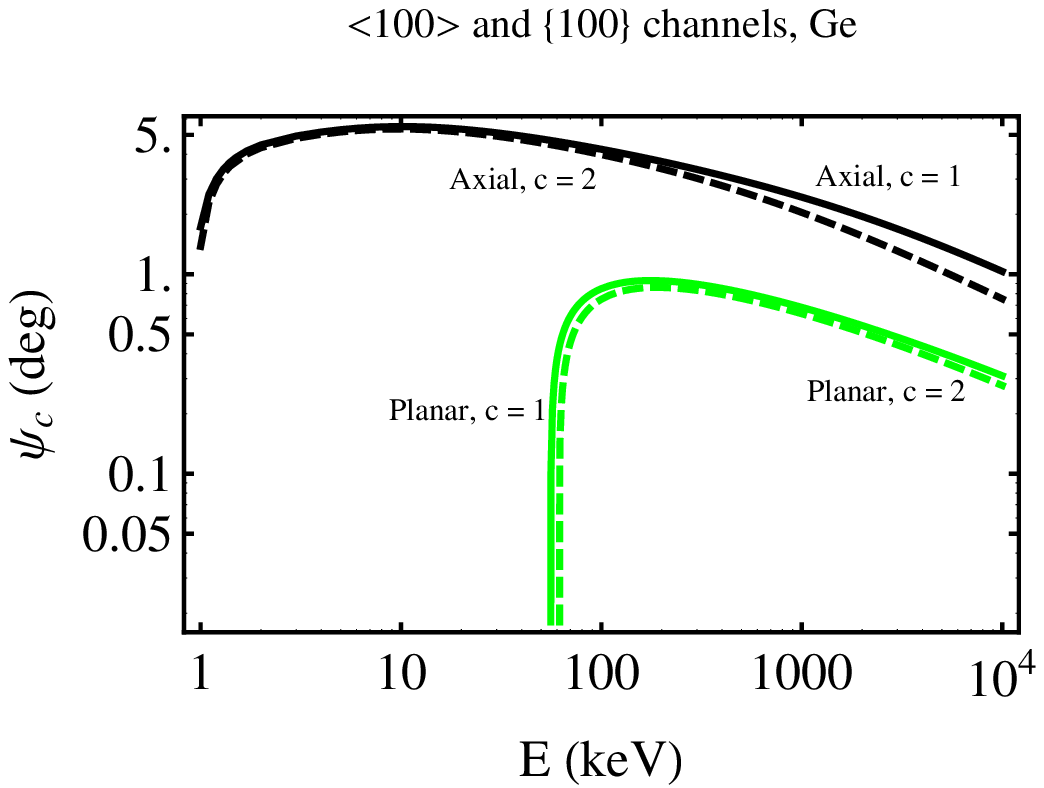,height=160pt}
        \caption{Same as in Fig.~\ref{rc-psic-MV-100-Si}  but for Ge ions propagating in
  the $<$100$>$ axial (black) and \{100\} planar (green/light gray) channels of a Ge crystal.}%
	\label{rc-psic-MV-100-Ge}}

 \FIGURE{\epsfig{file=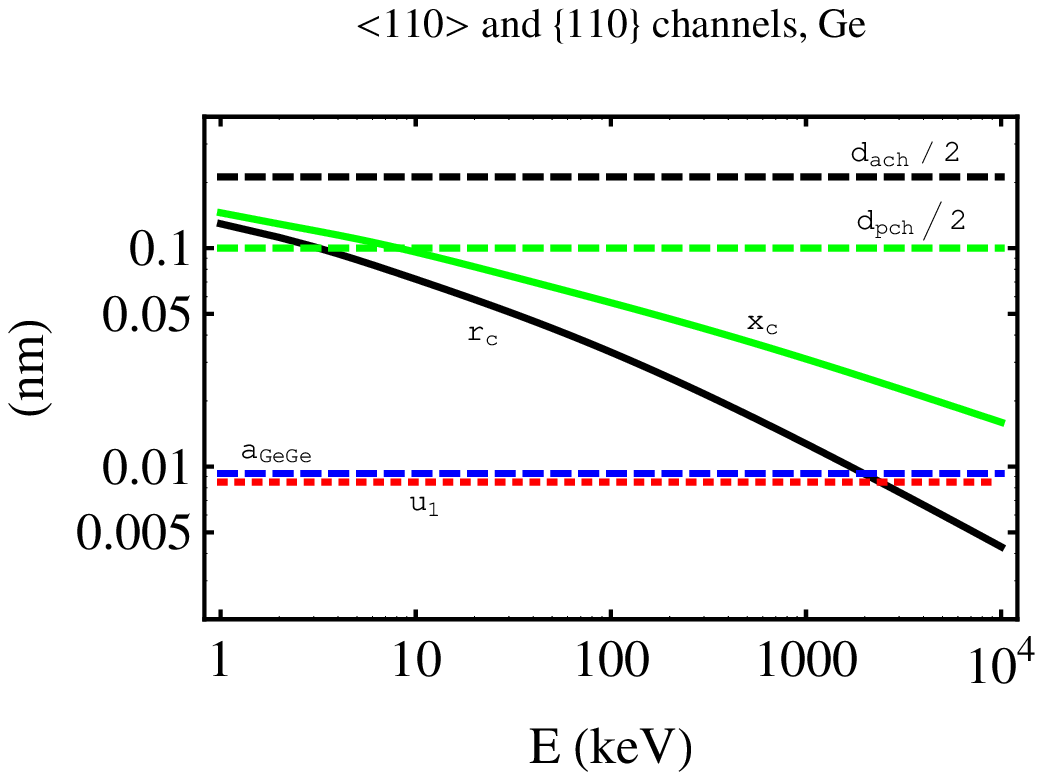,height=160pt}
 \epsfig{file=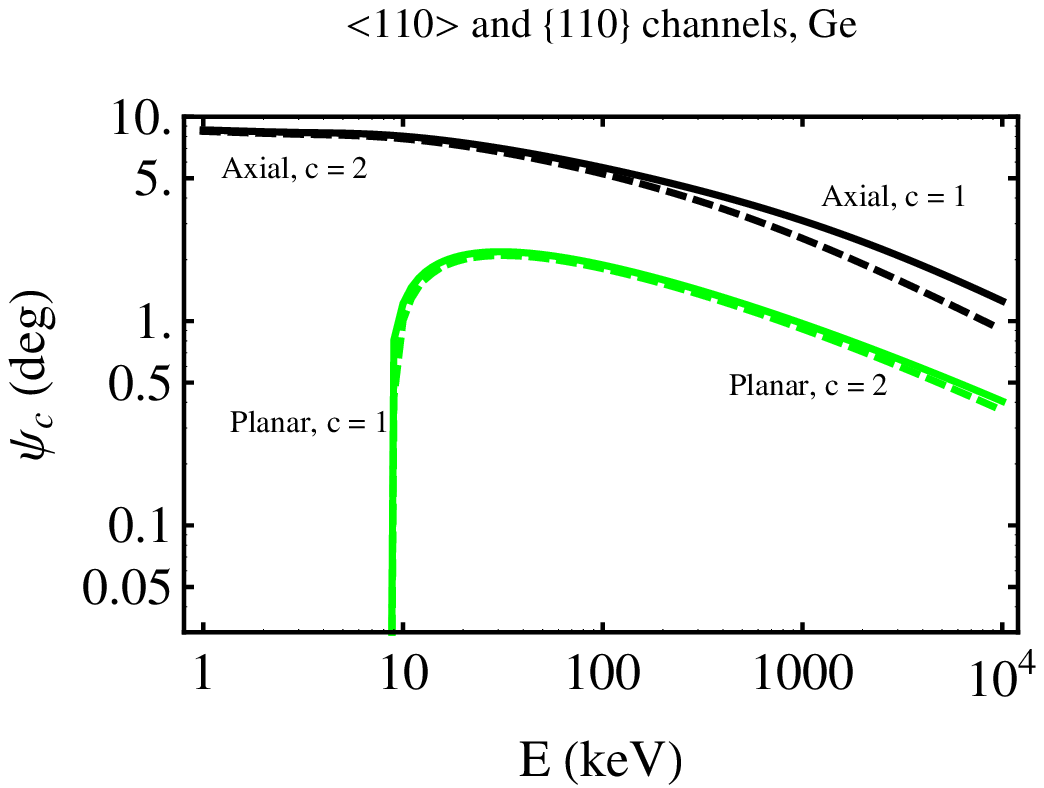,height=160pt}
        \caption{Same as in Fig.~\ref{rc-psic-MV-100-Ge} but for the $<$110$>$ axial and \{110\} planar channels.}%
	\label{rc-psic-MV-110-Ge}}
 \FIGURE{\epsfig{file=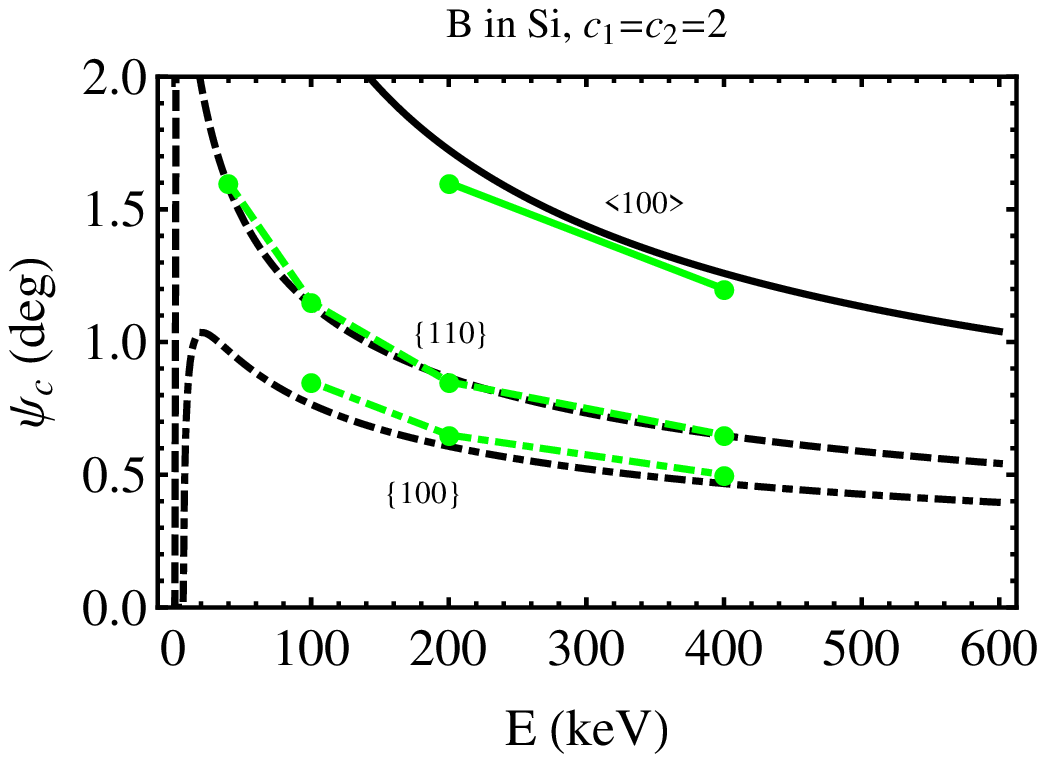,height=150pt}
 \epsfig{file=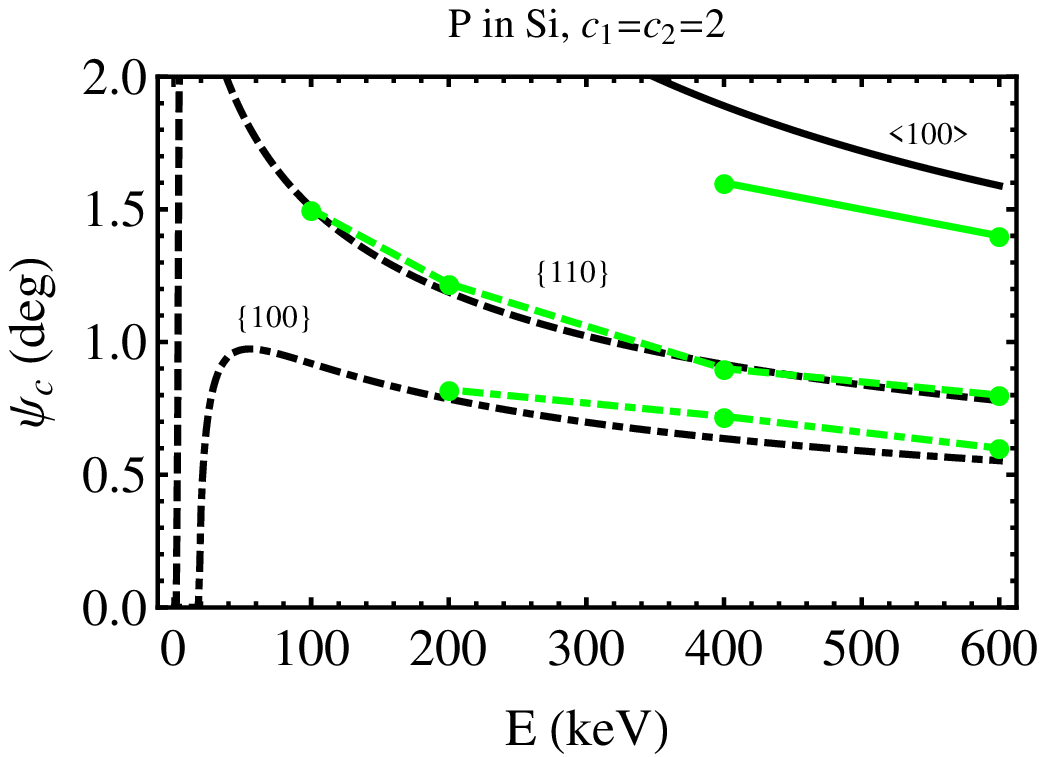,height=155pt}
        \caption{Comparison of theoretical (black lines) temperature corrected critical angles (with $c_1=c_2=2$) and measured critical angles at room temperature extracted from thermal wave measurements~\cite{Hobler} (green, or gray if color not available, dots  joined by straight lines to guide the eye) as a function of the energy of (a) B  ions and (b) P ions propagating  in a Si crystal at T=20 $^{\circ}$C, for the indicated axial and planar channels.}%
	\label{Compare-Data}}

In principle there are modifications to the continuum potentials due to thermal effects, but we are going to take into account thermal effects  in the crystal through a modification of the critical distances found originally by Morgan and Van Vliet~\cite{Morgan-VanVliet}  and later by Hobler~\cite{Hobler} to provide good agreement with simulations and data. For axial channels it consists of taking the temperature corrected critical distance $r_c(T)$  to be,
\begin{equation}
r_c(T)= \sqrt{r^2_c(E) + [c_1 u_1(T)]^2},
\label{rcofT}
\end{equation}
where the dimensionless factor $c_1$ in different references is a number between 1 and 2 (see  e.g. Eq. 2.32 of Ref.~\cite{VanVliet} and
Eq. 4.13 of the 1971 Ref.~\cite{Morgan-VanVliet}).

\FIGURE{\epsfig{file=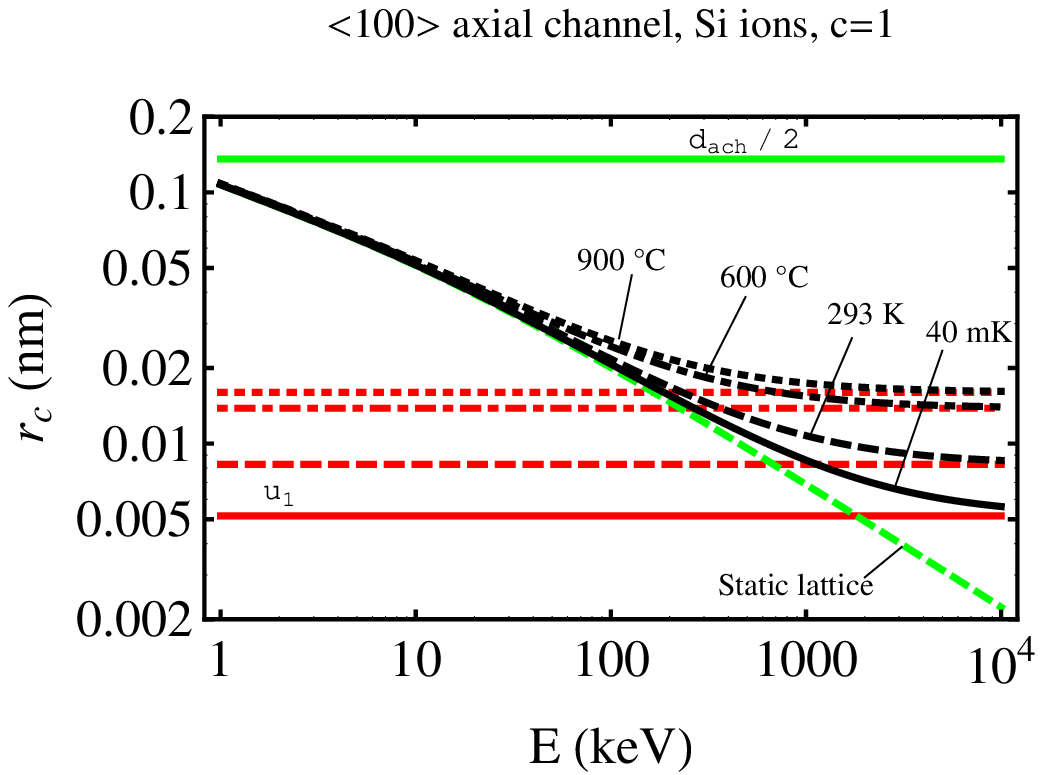,height=160pt}
 \epsfig{file=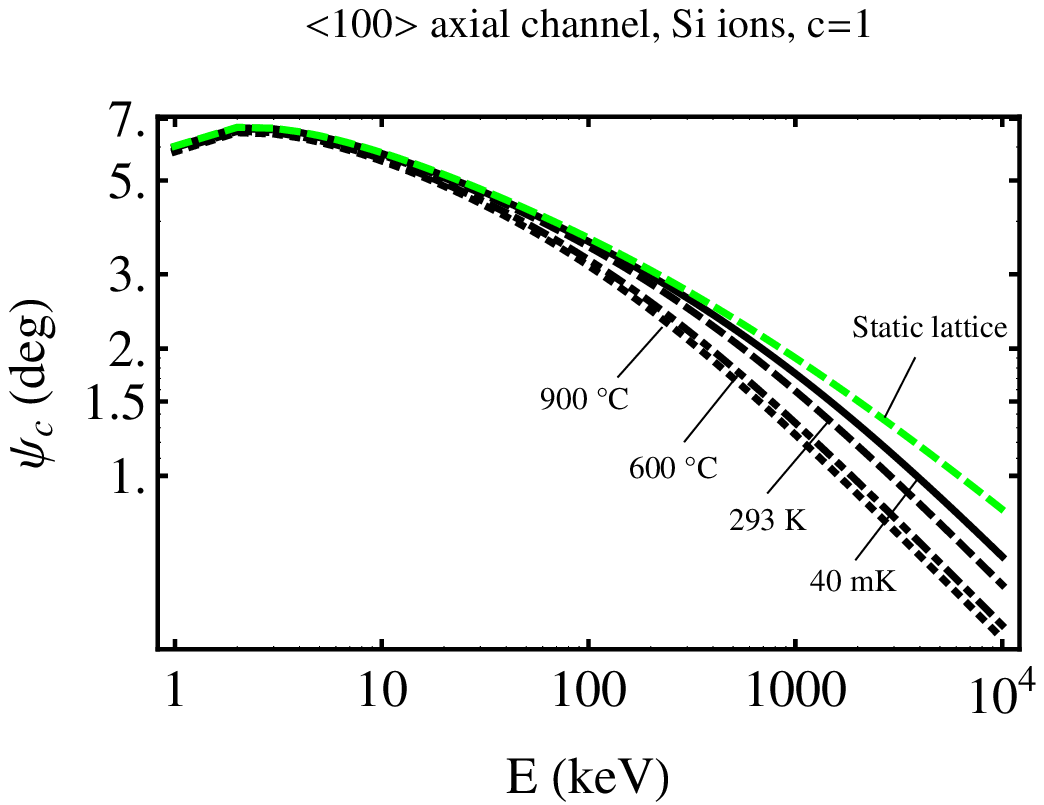,height=160pt}
        \caption{Static (green) and temperature corrected  with $c_1=c_2=c=1$ (black) (a) critical distances of approach (and  $u_1(T)$ in red) and (b) the corresponding critical channeling angles as a function of the energy of propagating  Si ions in $<$100$>$ axial  channels of a Si crystal.}%
	\label{rc-psic-MV-100-Si-c1}}
\FIGURE{\epsfig{file=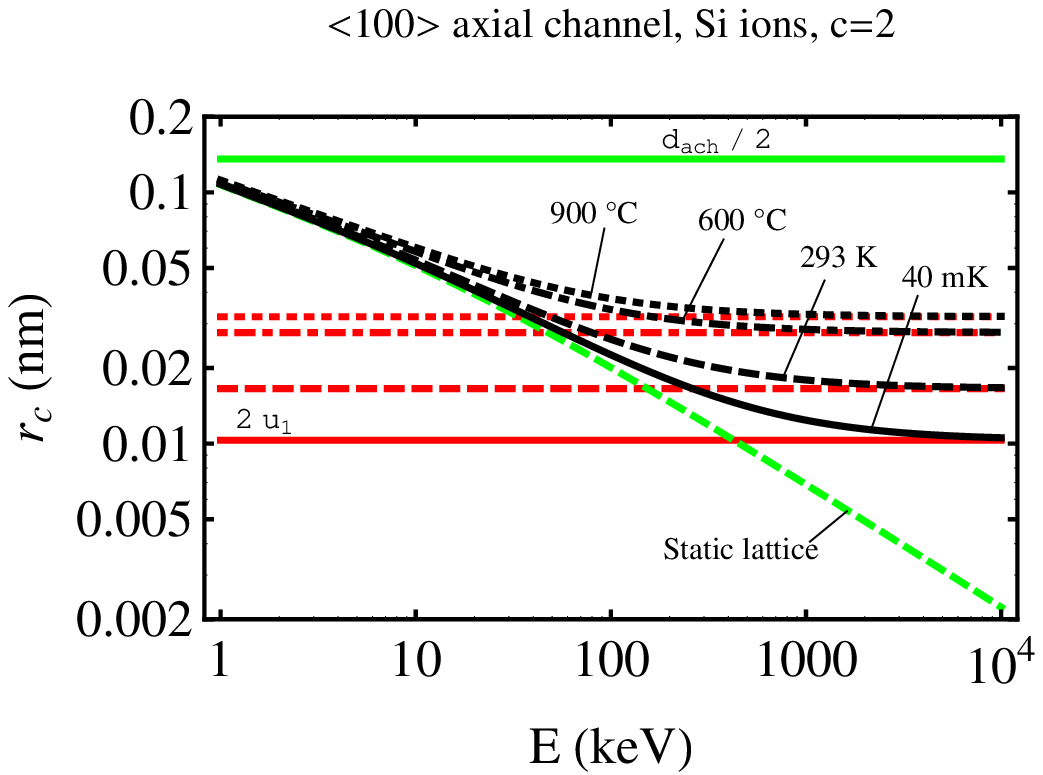,height=160pt}
 \epsfig{file=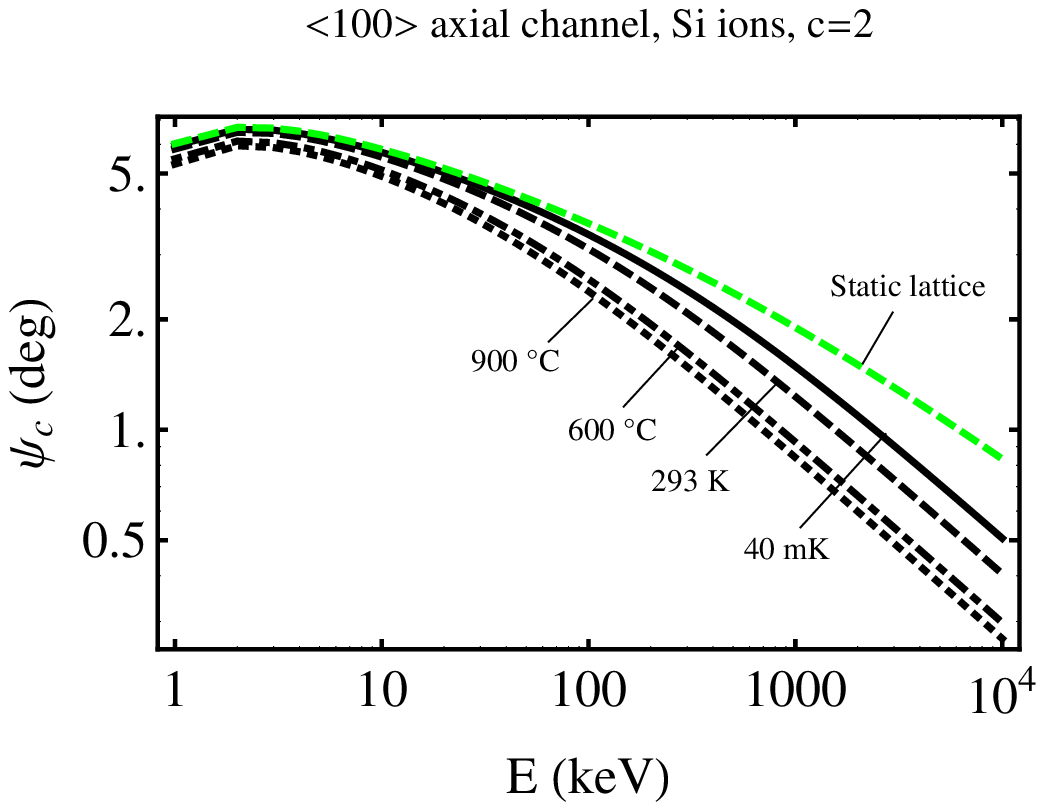,height=160pt}
        \caption{Same as Fig.~\ref{rc-psic-MV-100-Si-c1} but using $c_1=c_2=c=2$ in the temperature corrected critical distances of approach.}%
	\label{rc-psic-MV-100-Si-c2}}

For planar channels the situation is more complicated, because some references give a linear and others a quadratic relation between $x_c(T)$ and $u_1$. Following Hobler~\cite{Hobler} we use an equation similar to that for axial channels, namely
\begin{equation}
x_c(T)= \sqrt{x^2_c(E) + [c_2 u_1(T)]^2},
\label{xcofT}
\end{equation}
where again $c_2$ is a number between 1 and 2 (for example Barret~\cite{Barrett:1971} finds $c_2 = 1.6$ at high energies, and Hobler~\cite{Hobler} uses $c_2 = 2$). We will mostly use $c_1=c_2=1$ in the following, to try to produce upper bounds on the channeling fractions.

Using the temperature corrected critical distances of approach $r_c(T)$ and $x_c(T)$ (Eqs.~\ref{rcofT} and \ref{xcofT}) instead of the static lattice critical distances $r_c$ and $x_c$ (Eqs.~\ref{rcritPoly} and \ref{ourxcrit}),  in the definition of critical angles, Eqs.~\ref{psicritaxial} and \ref{psicritplanar}, we obtain the temperature corrected critical axial and planar angles, examples of which are shown in the right panels of Figs.~\ref{rc-psic-MV-100-Si} to \ref{rc-psic-MV-110-Ge} ($c_1=c_2=c$ and $c=1$ or  $c=2$ at room temperature).

As shown in Fig.~\ref{Compare-Data}, with this formalism and using $c_1=c_2=2$ we  fit relatively well the critical angles   measured  at room temperature for  B and P ions  in a Si  crystal (shown in green, or gray if color not available) in several channels, for energies between 20 keV and 600 keV that Hobler~\cite{Hobler} extracted from thermal wave measurements.

Figs.~\ref{rc-psic-MV-100-Si-c1} and \ref{rc-psic-MV-100-Si-c2} show clearly the temperature effects in the critical distances and angles for a specific channel, the  $<$100$>$ axial  channel of a Si crystal and for a propagating Si ion. At small energies the static critical distance of approach is much larger than the vibration amplitude, so temperature corrections are not important.  For small enough energies the critical distance becomes larger than the radius of the channel indicating that nowhere in the channel an ion can be far enough from the row of lattice atoms for channeling to take place  (thus the critical channeling angle is zero). The exact calculation of the energy at which this happens would require considering the effect of more than a single row of atoms (which we do not do here) thus our results at these low energies are only approximate.
As the energy increases, the static critical distance of approach decreases and when it becomes small with respect to  the vibration amplitude $u_1$, the temperature corrected critical distance becomes equal  to $(c_1 u_1)$ which is larger for larger  values of $c_1$. When $u_1(T)$ becomes important in determining the critical distance, this becomes larger, and therefore the critical channeling angle become smaller, for higher temperatures.

Figs.~\ref{T-depent-psic-Si} and \ref{T-depent-psic-Ge} show how the critical channeling angles change with temperature for four particular channels, the $<$110$>$ and $<$100$>$ axial and the \{110\} and \{100\} planar channels, for Si ions in Si and Ge ions in Ge, respectively. In both cases the axial channeling angles are larger  than the planar critical angles.  The  $<$110$>$
and \{111\} critical channeling angles are the largest among the axial and planar channels respectively.
 For example, at $E=200$ keV  for Si ions in Si, the channels with the largest channeling angles are  (in order of decreasing channeling angles): $<$110$>$, $<$100$>$, $<$211$>$, $<$111$>$, \{111\}, $<$311$>$, \{110\}, \{100\}, \{310\}, and \{210\}.
\FIGURE{\epsfig{file=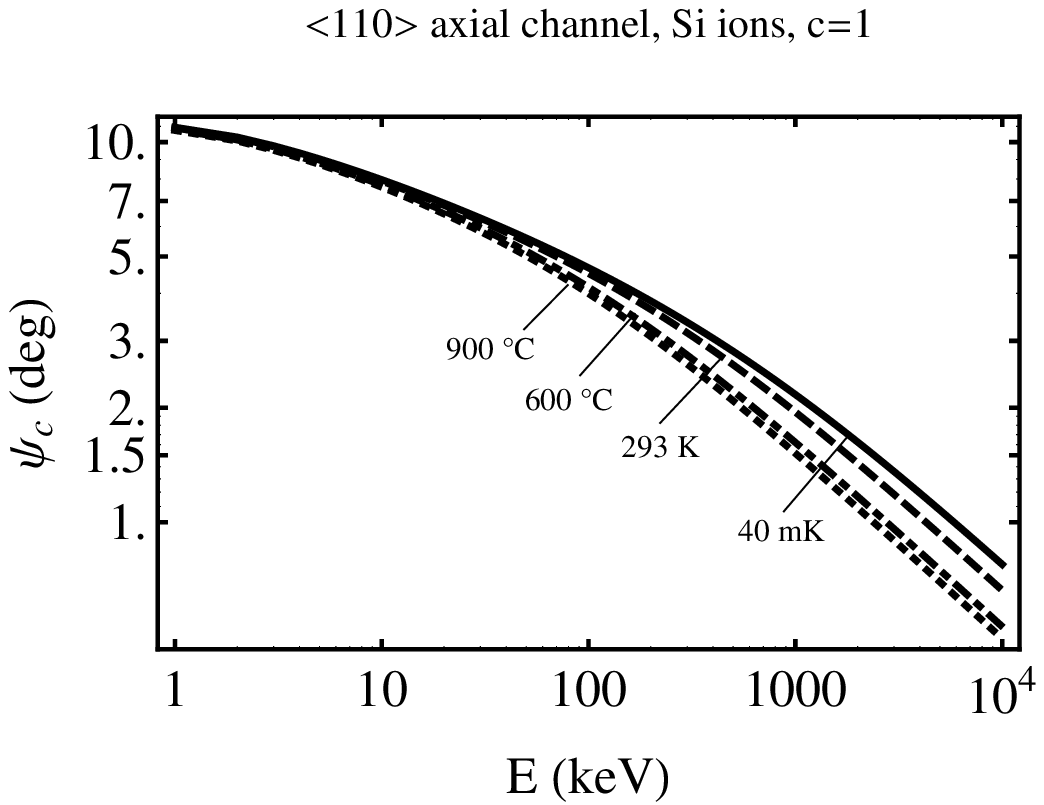,height=165pt}
 \epsfig{file=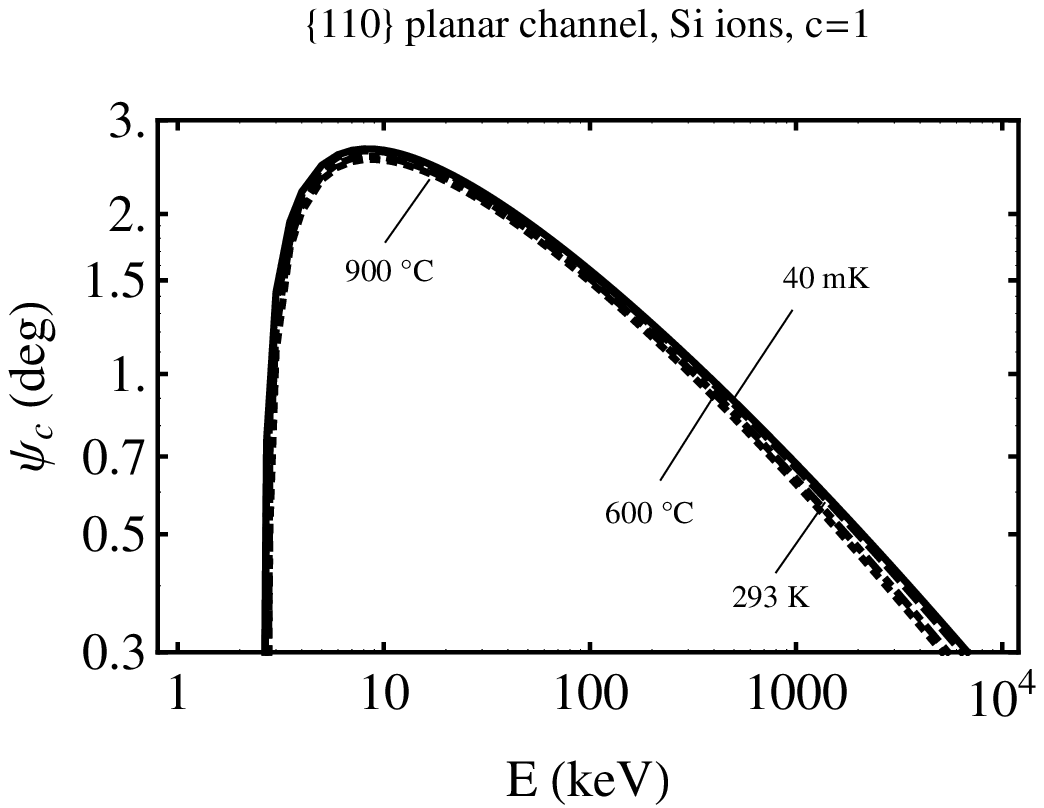,height=165pt}
        \caption{Temperature corrected critical channeling angles for T=40 mK, T=20 $^{\circ}$C, T=600 $^{\circ}$C, and T=900 $^{\circ}$C as a function of  the energy of a Si ion propagating in the (a) $<$110$>$ axial channels and (b) \{110\} planar channels of  Si crystal.}%
	\label{T-depent-psic-Si}}
\FIGURE{\epsfig{file=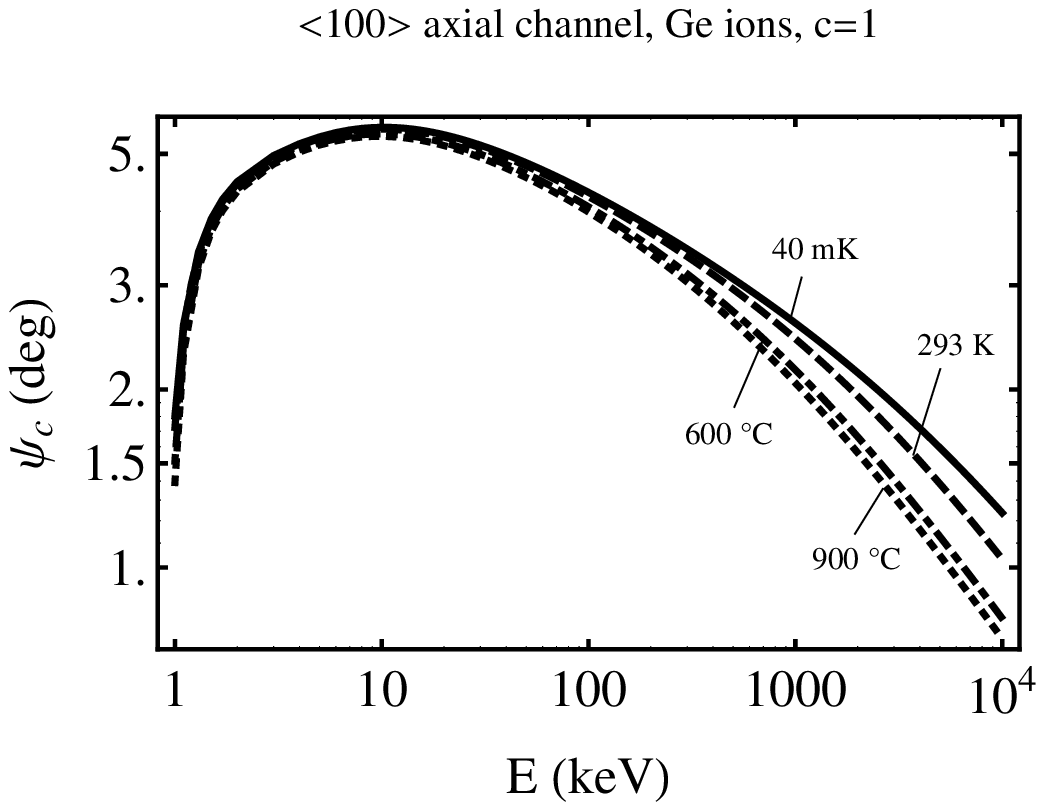,height=160pt}
 \epsfig{file=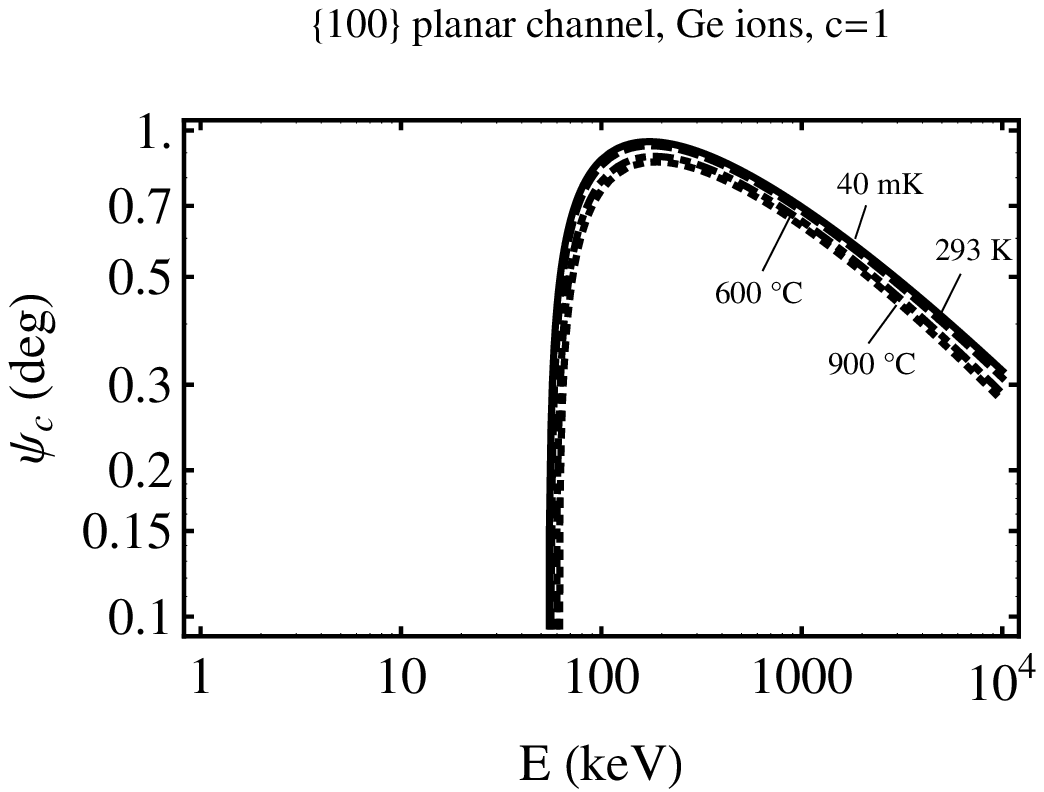,height=160pt}
        \caption{Same as Fig.~\ref{T-depent-psic-Si} but for a Ge ion propagating in the (a) $<$100$>$ axial channels and (b) \{100\} planar channels of a Ge crystal.}%
	\label{T-depent-psic-Ge}}
We can clearly see that the critical angles become zero at low enough energies (for which the critical distance of approach needed for channeling should be larger than the radius of the channel) indicating the range of energies for which no channeling is possible.

\section{Channeling of recoiling lattice nuclei}

 The channeling of ions in a crystal depends not only on the angle their initial trajectory makes with  rows or planes, but also on their initial position.
The nuclei  recoiling after an interaction with a WIMP start initially from lattice sites (or very close to them), thus blocking effects  are important. In fact, as argued originally by Lindhard~\cite{Lindhard:1965}, in a perfect lattice and in the absence of energy-loss processes the probability of a particle starting from a lattice site to be channeled would be zero.
This is what Lindhard called the {\it ``Rule of Reversibility."}
 However, any departure of  the actual lattice from a perfect lattice, for example due to vibrations of the atoms in the lattice, violates the conditions of this argument and allows for some of the recoiling lattice nuclei to be channeled. Lattice vibrations are more important at hight temperatures and they have two opposite effects on channeling fractions: the probability of finding the atom which collides with a dark matter particle further out of its equilibrium position increases with increasing temperature  thus channeling fractions increase, but the range of angles the trajectory of the propagating ion must make with the direction of the channel decreases with increasing temperature, which decreases the channeling fraction.

   We now estimate the channeling fraction using the formalism presented so far.

\subsection{Channeling fraction for each channel}

As in Ref.~\cite{BGG-I} we use a Gaussian function for the probability distribution
$g(r)$ for the perpendicular distance $r$  to the row at which  the atom  that collides with a WIMP is located at the moment of the collision due to thermal vibrations in the crystal
\begin{equation}
g(r)=\frac{r}{u_1^2}\exp{(-r^2/2u_1^2)}.
\label{gofr}
\end{equation}
The one dimensional vibration amplitude $u_1$ is given in Eq.~\ref{vibu1}. As explained in detail in  Ref.~\cite{BGG-I}, the channeled fraction $\chi_{\rm axial}(E, \hat{\textbf{q}})$ of nuclei with recoil energy $E$ moving initially in the direction $\hat{\textbf{q}}$ making an  angle $\phi$ with respect to an atomic row is given by the fraction of nuclei  which can be found at a distance $r$ larger than a minimum distance $r_{i,\rm min}$ from the row at the moment of collision, determined by the critical distance of approach
\begin{equation}
\chi_{\rm axial}(E, \phi)=\int_{r_{i,\rm min}}^{\infty}{dr g(r)}=\exp{(-r_{i,\rm min}^2/2u_1^2)}.
\label{chiaxial}
\end{equation}
It can easily be seen using  Eqs.~\ref{E-perp-HP},  \ref{r*definition},  \ref{eq:consetrans}  and  \ref{defrcrit} that, because $U(r_i + d \tan\phi) \geq U(r_c)$,  if $\phi>\psi_c$ no channeling can occur and $\chi_{\rm axial}(E, \phi)=0$.

Using the condition
\begin{equation}
E \sin^2 \phi + U(r_i + d \tan\phi)= U(r_{\rm min}) < U(r_c(E)),
\label{ChanCond-CA}
\end{equation}
that implies the equality for the minimum initial distance $r_{i,\rm min}$,
\begin{equation}
U(r_{i,\rm min}+ d \tan\phi) = U(r_c(E))-E \sin^2\phi,
\end{equation}
in Ref.~\cite{BGG-I} we derived the following analytic expression  for $r_{i,\rm min}$ from Lindhard's approximation to the potential:
\begin{equation}
r_{i,\rm min} (E, \phi) = \frac{C a}{\sqrt{\left( 1+\frac{C^2a^2}{r_{c}^2} \right) \, \exp\!\left(-{2 \sin^2\phi}/{\psi_1^2} \right) -1}}
- d \tan\phi,
\label{rimin}
\end{equation}
where $C$ is a constant, which was found experimentally to be $C\simeq\sqrt{3}$~\cite{Lindhard:1965}. We use here this  equation because it is not possible to find a similar analytic expression using Moli\`{e}re's approximation to the potential (although following Hobler we use  Moli\`{e}re's approximation to obtain the critical distances and angles). $r_{i,\rm min}$ is a function of the temperature too, through $r_{c}(T)$. Notice that a small change in the critical distance $r_c(T)$ and thus in $r_{i,\rm min}$ is exponentially magnified in the channeling fraction $\chi_{\rm axial}$ (Eq.~\ref{chiaxial}). This constitutes the most important difficulty to   evaluate channeling fractions. The same happens for planar channels.

For a planar channel, the Gaussian thermal distribution for the planar potential is one-dimensional (the relevant vibrations occurring perpendicularly to the plane),
\begin{equation}
  g(x)=  (2 \pi u_1^2)^{-1/2} \exp(-x^2/2u_1^2).
\label{gofx}
\end{equation}
This is normalized  to  be 1 for $-\infty<x<+\infty$. In our calculations we only consider positive values of $x$,  $0<x<+\infty$, for each plane, thus we multiply $g(x)$ by a factor of $2$ to find the fraction of channeled nuclei for a planar channel,
\begin{equation}
\chi_{\rm planar}(E, \phi)=\int_{x_{i,\rm min}}^{\infty}{2~g(x) dx}
=\frac{2}{\sqrt{\pi}}\int _{x_{i,\rm min}}^{\infty}{\frac{\exp(-x^2/2u_1^2)}{\sqrt{2}u_1}dx}\nonumber\\
=\mathop{\rm erfc}\left(\frac{x_{i,\rm min}}{\sqrt{2}u_1}\right),
\label{chiplanar}
\end{equation}
where the minimum initial distance (derived in Ref.~\cite{BGG-I} using Lindhard's planar potential) is
\begin{equation}
x_{i, \rm min}(E, \phi)= \frac{(a/2) \left\{C^2-\left[\sqrt{(x_c^2/a^2)+C^2}- (x_c/a)- (\sin^2\phi/\psi_a^2)\right]^2\right\} }{\left[\sqrt{(x_c^2/a^2)+C^2}-(x_c/a)-(\sin^2\phi/\psi_a^2)\right]} - d_p \tan\phi.
\label{ximin}
\end{equation}
Here $\phi$ is the angle $\hat{\textbf{q}}$ makes with the plane, defined as the complementary angle to the angle between $\hat{\textbf{q}}$ and the normal to the plane, or as the smallest angle between $\hat{\textbf{q}}$ and vectors lying on the plane. Also in this case, $\chi_{\rm planar}(E, \phi)=0$ if $\phi>\psi_c^p$. Note that $x_{i, \rm min}$ is also a function of the temperature through its dependence on  $x_c(T)$.

Using either Eq.~\ref{chiaxial} or Eq.~\ref{chiplanar}, for an axial and a planar channel respectively, we define
channeling fraction $\chi_k$ for each channel $k$, which depends on
 the initial energy $E$, initial angle $\phi$ and temperature $T$. Then we will sum over all channels  and angles to obtain the total channeling fraction as function of $E$ and $T$.
\FIGURE{\epsfig{file=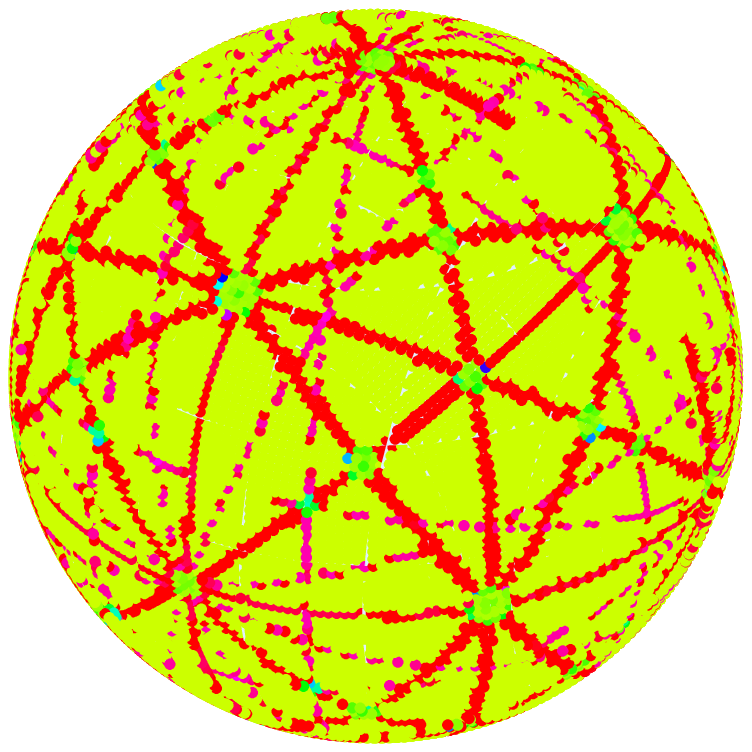,height=170pt}
 \epsfig{file=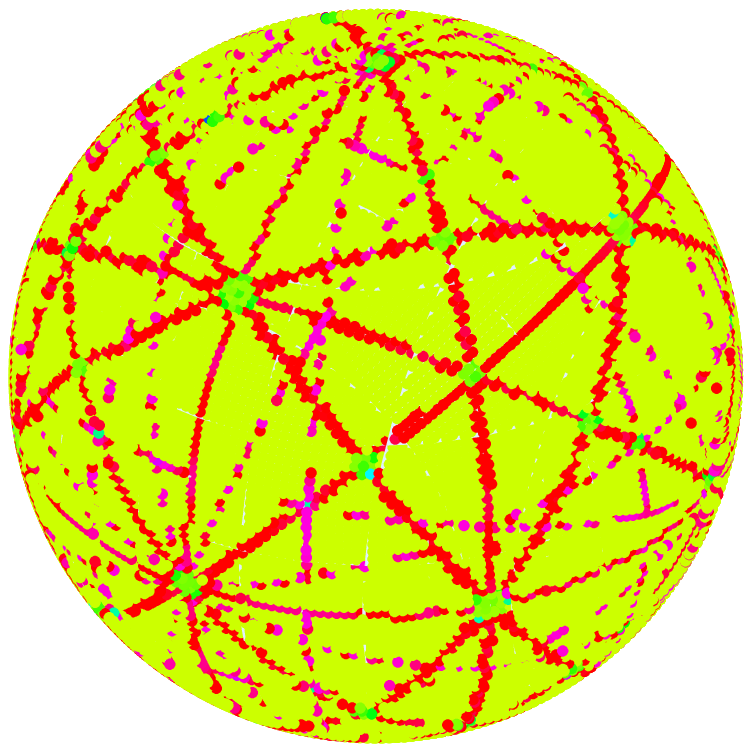,height=170pt}
       \vspace{-0.5cm} \caption{Geometric channeling fraction $\chi_{\rm rec}(E, {\bf{\hat{q}}})$  including 74 channels  (Eq.~\ref{chirec}) for each direction ${\bf{\hat{q}}}$ plotted using the HEALPix pixelization of a sphere, for (a) a 200 keV Si  ion recoil in a Si crystal, and (b) a 1 MeV Ge ion recoil in a Ge crystal, at 20 $^o$C. Temperature  effects in the lattice were included with  $c_1=c_2=1$. The light green, light blue, dark blue, pink, red, and yellow colors indicate a channeling fraction of 0.5, 0.013, $7.5 \times 10^{-4}$, $4 \times 10^{-5}$, $10^{-5}$ and zero, respectively.}%
	\label{Our-HEALPIX}}

\subsection{Total geometric channeling fraction}

The geometric channeling fraction is the fraction of recoiling ions that propagate in the 1st, or 2nd, or \ldots or 74th channel. Here ``geometric'' refers to assuming that the distribution of recoil directions is isotropic. In reality, in a dark matter direct detection experiment, the distribution of recoil directions is expected to be peaked in the direction of the average WIMP flow. Here we examine this geometrical channeling fraction, and postpone the case of a WIMP wind to another paper \cite{NGG-WIMPwind}.

We include in our calculation only the most important channels, the same considered by Hobler~\cite{Hobler}. These are  the  $<$100$>$, $<$110$>$, $<$111$>$, $<$211$>$ and $<$311$>$ axial channels and the \{100\}, \{110\}, \{111\}, \{210\} and \{310\} planar channels. These constitute a total of 74 channels, as explained in Appendix B.

The probability $\chi_{\rm rec}(E,\hat{\bf q})$ that an ion with initial energy $E$ is channeled in a given direction $\hat{\bf q}$ is the probability that the recoiling ion enters any of the available channels, i.e.
\begin{equation}
\chi_{\rm rec}(E,\hat{\bf q}) = P(A_1~\text{or}A_2~\text{or \ldots or}~A_{74}).
\label{chirec}
\end{equation}
We compute this probability in the same way we did in Ref.~\cite{BGG-I}, using a recursion of the addition rule in probability theory
and treating channeling along different channels as independent (see in Appendix C that this is a good approximation).

 Fig.~\ref{Our-HEALPIX} shows the channeling probability $\chi_{\rm rec}(E, {\bf{\hat{q}}})$  in Eq.~\ref{chirec} for a 200 keV recoiling Si ion  in  a Si crystal and a 1 MeV Ge ion recoil in a Ge crystal, at 20 $^\circ$C. Temperature effects were included with $c_1=c_2=1$.
  The probability is computed for each direction and plotted using the HEALPix pixelization of a sphere. The light green, light blue, dark blue, pink, red, and yellow colors indicate a channeling probability of 0.5, 0.013, $7.5 \times 10^{-4}$, $4 \times 10^{-5}$, $10^{-5}$ and zero, respectively.

To obtain the geometric total channeling fraction, we average the channeling probability $\chi_{\rm rec}(E,\hat{\bf q})$ over the directions $\hat{\bf q}$, assuming an isotropic distribution of the initial recoiling directions $\hat{\bf q}$,
\begin{equation}
P_{\rm rec}(E)=\frac{1}{4\pi}\int{\chi_{\rm rec}(E, \hat{\bf q})d\Omega_q}.
\end{equation}
This integral is computed using HEALPix~\cite{HEALPix:2005} (see Appendix B of Ref.~\cite{BGG-I} for a complete explanation).

Our results for the geometric total channeling fraction for Si ions in a Si crystal  and  Ge ions in a Ge crystal are shown in Figs.~\ref{FracSiG-DiffT-rigid}, \ref{FracSiG-DiffT-c1} and \ref{FracSiG-DiffT-c2} for three different assumptions for the effect of thermal vibrations in the lattice, which depend on the values
of the parameters $c_1$ and $c_2$ used in  the temperature corrected critical distances of approach $r_c(T)$ and $x_c(T)$ in Eqs.~\ref{rcofT} and \ref{xcofT}. The unrealistic case of assuming no vibrations in the lattice (except for vibrations of the colliding atom) corresponds to taking $c_1=c_2=0$ and is shown in Fig~\ref{FracSiG-DiffT-rigid} for different temperatures because it provides  an upper limit to the channeling fractions. In this case the channeling fractions reach a few \% and they increase with temperature.

  In the literature, in other materials or for other channeling ions, values of $c_1$ and $c_2$ between 1 and 2 are used. Thus, we show the $c_1=c_2=1$ choice in Fig.~\ref{FracSiG-DiffT-c1} and the $c_1=c_2=2$ in Fig.~\ref{FracSiG-DiffT-c2}. As the values  of $c_1$ and $c_2$ increase,  also the minimal distances from row or planes at which propagating ions must be to be channeled increase, thus  the critical channeling angles decrease, what makes the channeling fractions smaller. If the values of $c_1$ and $c_2$ found by Hobler~\cite{Hobler} and by us (see Fig.~\ref{Compare-Data}) to fit measured channeling angles for B and P ions propagating in Si apply also to the propagation of Si ions in Si, then the case of  $c_1=c_2=2$ in Fig.~\ref{FracSiG-DiffT-c2} should be chosen and the channeling fractions would never be larger than 0.3\%. With  $c_1=c_2=1$ the channeling fractions reach  about 1\% and they increase with temperature.

Please note that we have not considered the possibility of dechanneling of initially channeled ions due to imperfections in the crystal. Any mechanism of dechanneling will decrease the fractions obtained here.
\FIGURE[t]{\epsfig{file=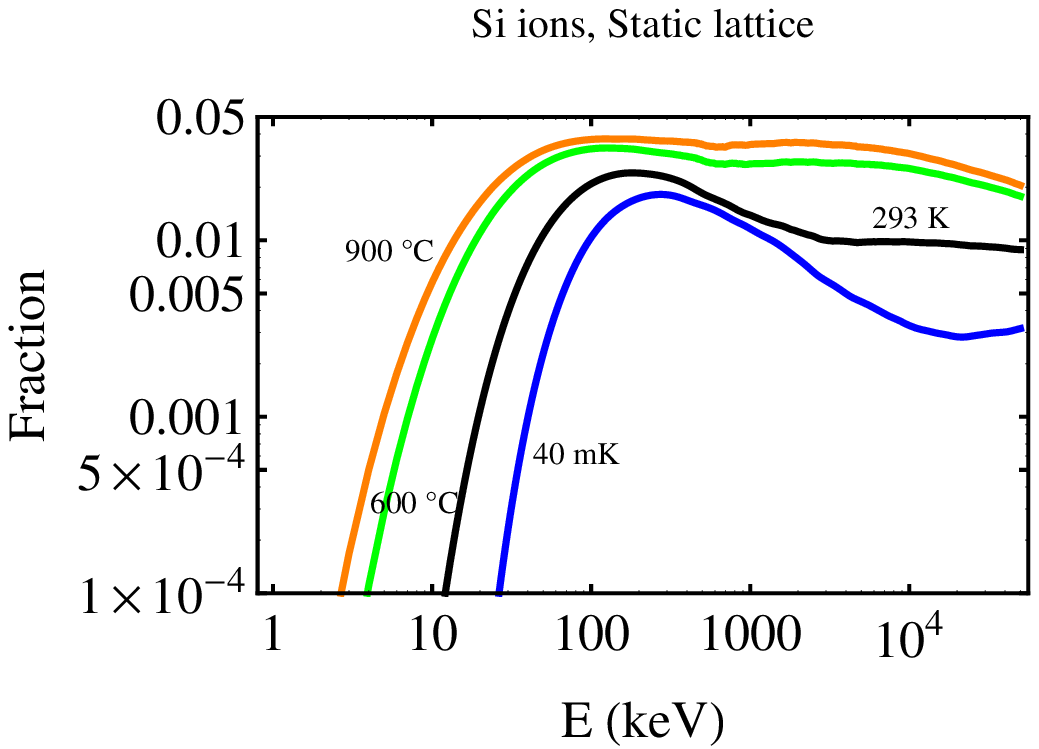,height=150pt}
 \epsfig{file=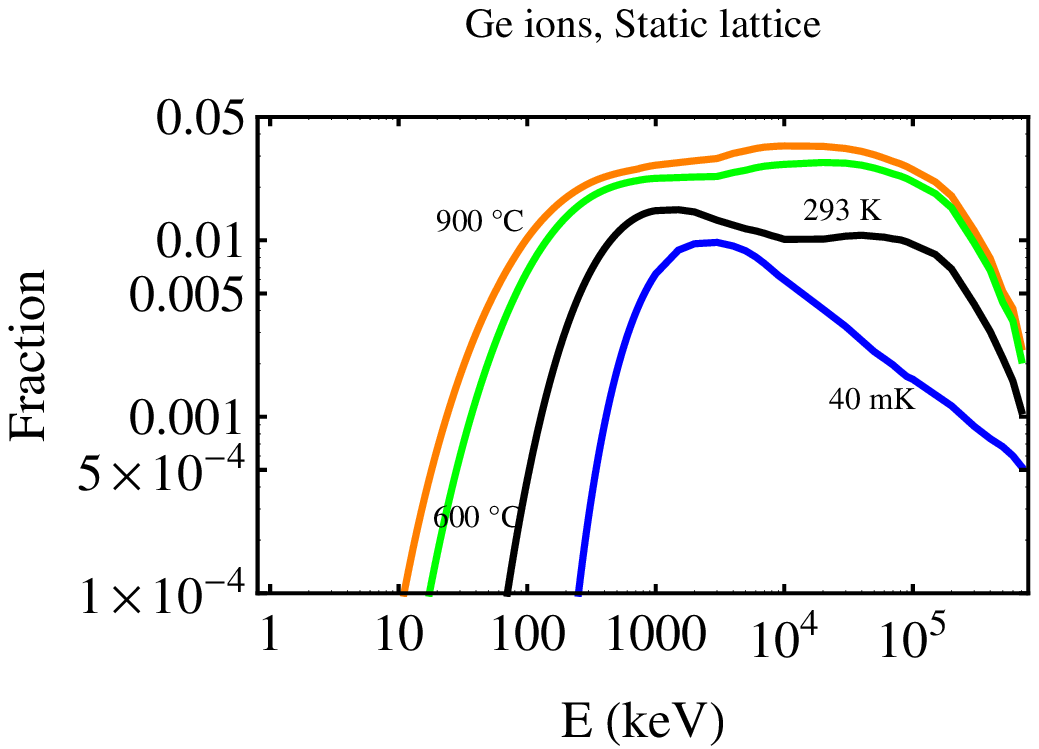,height=150pt}
      \caption{Channeling fractions of (a) Si  and (b) Ge recoils in a Si and a Ge crystal respectively, as a function of the ion energy for  temperatures T=900 $^\circ$C (orange or medium gray), 600 $^\circ$C (green or light gray), 293 K (black), and 44 mK (blue or dark gray)  in the approximation of $c_1=c_2=0$ (``static lattice"). This is  an upper bound with respect to any non-zero values of $c_1$ and $c_2$. Temperature effect are included in the vibrations of the colliding atom.}%
	\label{FracSiG-DiffT-rigid}}
\FIGURE{\epsfig{file=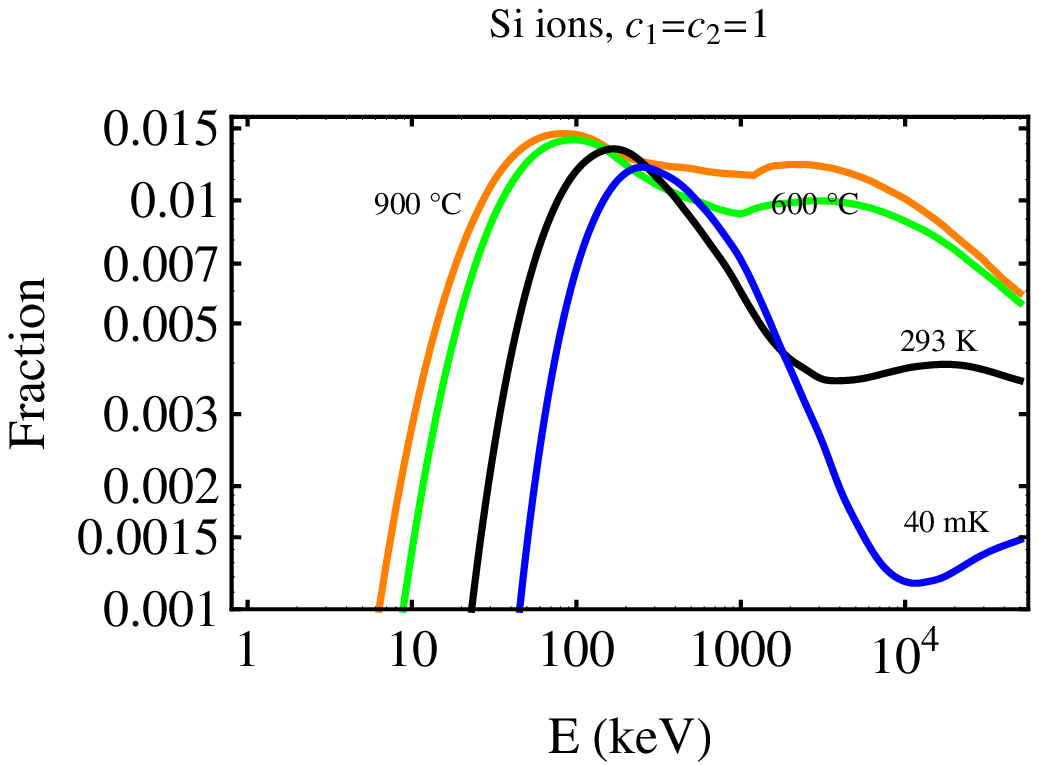,height=150pt}
 \epsfig{file=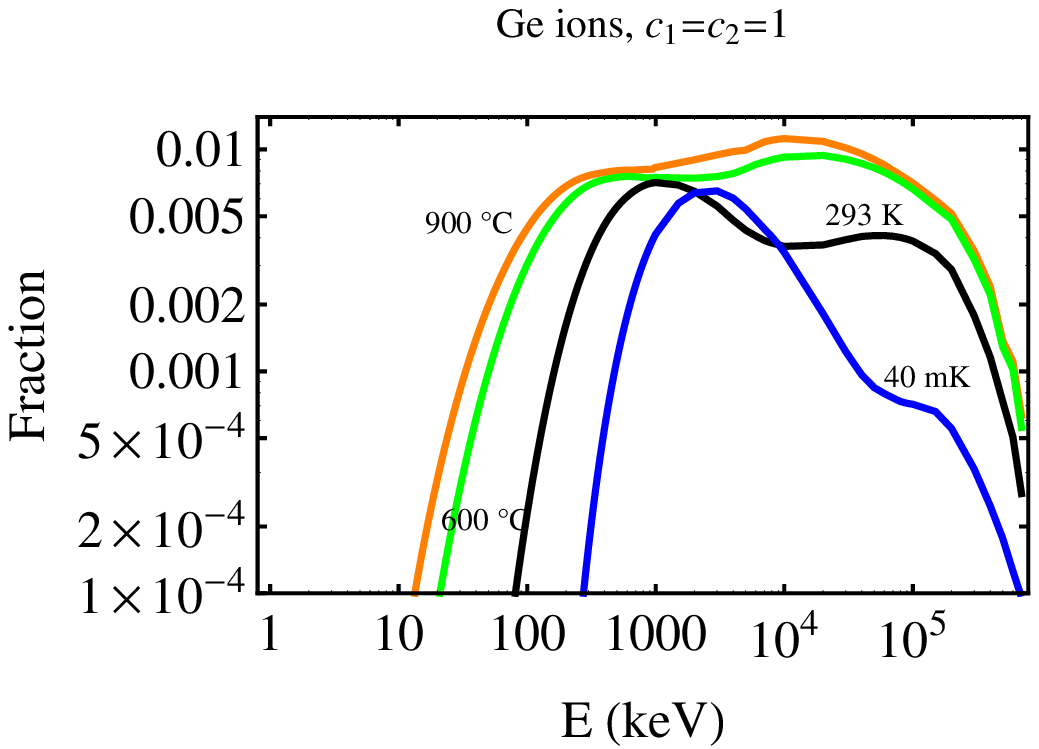,height=150pt}
      \caption{Same as Fig.~\ref{FracSiG-DiffT-rigid} but with $c_1=c_2=1$.}%
	\label{FracSiG-DiffT-c1}}
\FIGURE{\epsfig{file=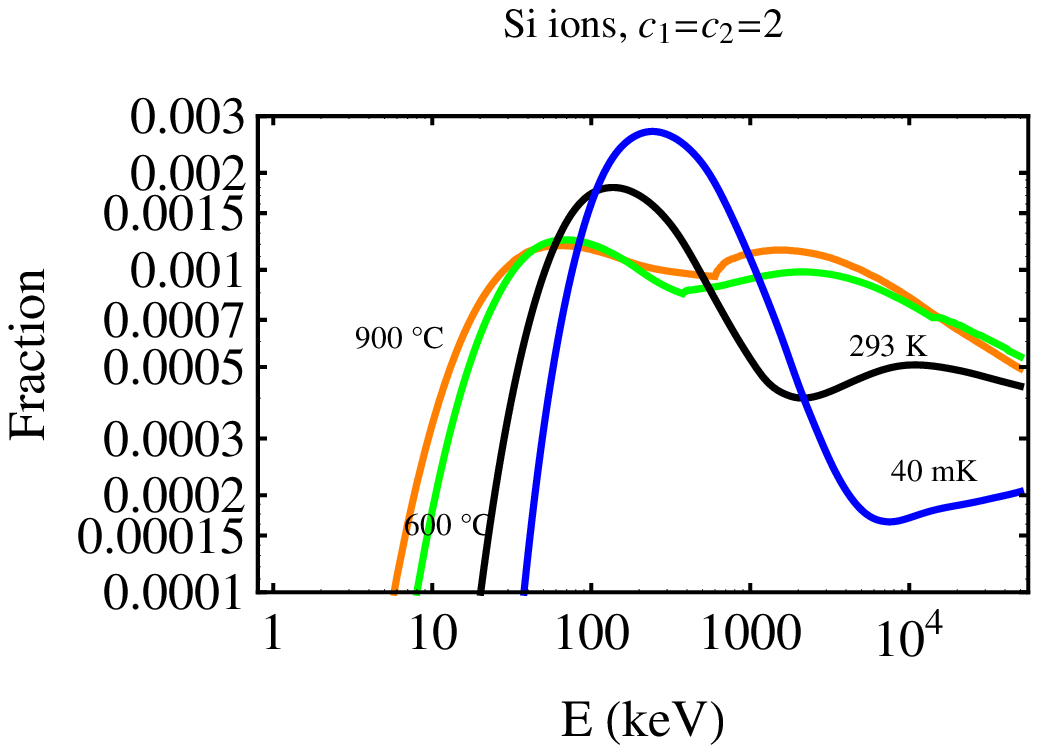,height=150pt}
 \epsfig{file=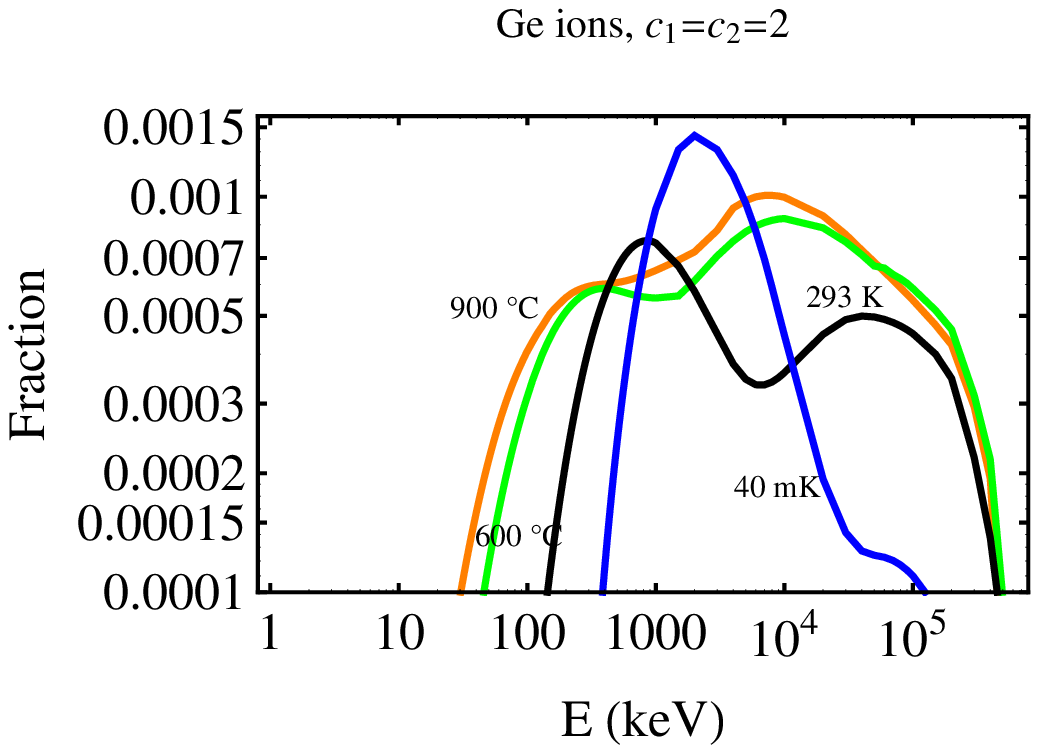,height=150pt}
      \caption{Same as Fig.~\ref{FracSiG-DiffT-rigid} but with $c_1=c_2=2$.}%
	\label{FracSiG-DiffT-c2}}

\section{Main results and conclusions}

We have studied the channeling of ions recoiling after a collision with dark matter particles within  Si and Ge crystals.   The calculations are similar because both crystals have the same structure. Channeled ions move within the crystal along symmetry axes and planes and suffer a series of small-angle scatterings  that maintain them in the open ``channels"  in between the rows or planes of lattice atoms and thus penetrate much further into the crystal than in other directions.  In order for the scattering to happen at small enough angles, the propagating ion must not approach a row or plane closer than a critical distance $r_c$ or $x_c$ respectively. These are given in Eqs.~\ref{rcritPoly}
and \ref{ourxcrit} for a ``static lattice" (i.e. a perfect lattice in which all vibrations are neglected) and by Eqs.~\ref{rcofT} and \ref{xcofT} once  temperature vibrations of the crystal lattice are taken into account. The temperature corrected minimum distances of approach (in Eqs.~\ref{rcofT} and \ref{xcofT})  depend on the one dimensional rms vibration amplitude $u_1(T)$ (Eq.~\ref{vibu1}), which increases with the temperature, through the coefficients $c_1$ and $c_2$. These dimensionless coefficients are found in the literature (for different ions and/or crystals) to take values between 1 and 2. Channeled ions must have trajectories that  at large distances from the atomic rows or planes must make an angle with respect to these rows or planes smaller than a critical angle given in Eqs.~\ref{psicritaxial} and \ref{psicritplanar} for axial and planar channels respectively.

 The critical  angles depend on the temperature through the minimum distances of approach: as these increase with increasing temperatures, the critical angles decrease, what makes the channeling fraction smaller. However,
there is a second temperature effect which makes the channeling fractions larger  as the temperature increases: the vibrations of the atom which collides with the dark matter particle. Thus, the channeling fraction of recoiling ions is strongly temperature dependent. Depending on which of the two competing effects is dominant, the channeling fraction may either increase or decrease as the temperature increases. Increasing the temperature of a crystal usually increases the fraction of channeled recoiling ions (see Figs.~\ref{FracSiG-DiffT-rigid} and \ref{FracSiG-DiffT-c1}), but when the values of $c_1$ and $c_2$ are large  (i.e. close to 2) so the critical distances increase rapidly with the temperature, the opposite may happen (see Fig.~\ref{FracSiG-DiffT-c2}).

The vibrations of the atom colliding with the dark matter particle are essential to have a non-zero probability of channeling of the recoiling ion. A nucleus ejected from its lattice site by a collision with a dark matter particle is initially part of a row or plane.  Thus, the recoiling nuclei start initially from lattice sites or very close to them. This means that blocking effects  are important. In fact, as argued originally by Lindhard~\cite{Lindhard:1965}, in a perfect lattice and in the absence of energy-loss processes the probability of a particle starting from a lattice site to be channeled would be zero.  This is what Lindhard called the {\it ``Rule of Reversibility."} However,  vibrations of the atoms in the lattice violate the conditions of this argument and allow for some of the recoiling lattice nuclei to be channeled.  The channeling fraction $\chi_{\rm axial}$, Eq.~\ref{chiaxial}, or $\chi_{\rm planar}$, Eq.~\ref{chiplanar} for axial and planar channels respectively, is given by the fraction of nuclei which can be found further than a minimum distance $r_{i,\rm min}$ or $x_{i,\rm min}$  away from a row or plane at the moment of collision. This fraction increases as $u_1(T)$ increases. This is the effect that dominates the temperature dependence in Figs.~\ref{FracSiG-DiffT-rigid} and \ref{FracSiG-DiffT-c1}, in which the geometric channeling fractions increase with increasing temperature.

  However, $r_{i,\rm min}$, Eq.~\ref{rimin} or $x_{i,\rm min}$, Eq.~\ref{ximin}, increase with increasing critical distances and this decreases the channeling fraction. The increase of the critical distances with temperature is more accentuated for large values of $c_1$ and $c_2$. This can be seen  in Fig.~\ref{FracSiG-DiffT-c2}, in which $c_1=c_2=2$ and some channeling fractions are larger at lower temperatures.

The unrealistic case of assuming no vibrations in the lattice (except for vibrations of the colliding atoms) corresponds to taking $c_1=c_2=0$ (this is what we call the ``static lattice" approximation) shown in Fig.~\ref{FracSiG-DiffT-rigid} for different temperatures, provides  an upper limit to the channeling fractions. This is the limiting case of the possibility that $c_1$ and $c_2$ are smaller than 1, in which  the channeling fractions reach a few \% at energies of 100's of keV and  increase with temperature.

  We show the $c_1=c_2=1$ choice in Fig.~\ref{FracSiG-DiffT-c1} and the $c_1=c_2=2$ in Fig.~\ref{FracSiG-DiffT-c2}.  If the values found by Hobler~\cite{Hobler} and by us (see Fig.~\ref{Compare-Data}) to fit the measured channeling angles for B and P ions propagating in a Si  crystal apply also to the propagation of Si ions in Si, then the case of
 $c_1=c_2=2$ should be chosen and the channeling fractions would never be larger than  a few 0.1\%. In this case, as mentioned above, the channeling fractions at some energies are higher at lower temperatures. The $c_1=c_2=1$ case, instead,  leads to maximum channeling fractions of roughly  1 \%, which increase with increasing temperature.

 Notice that a small change in the critical distances $r_c(T)$ or $x_c(T)$ and thus in the initial minimum distances  of approach $r_{i,\rm min}$ or $x_{i,\rm min}$ is exponentially magnified in the channeling fractions $\chi_{\rm axial}$, Eq.~\ref{chiaxial}, or $\chi_{\rm planar}$, Eq.~\ref{chiplanar}. This constitutes the most important difficulty to  evaluate channeling fractions in the models we use.

 Notice too that we have not considered any mechanism of dechanneling of the channeled ions (due to irregularities in the crystals, for example) which would decrease the channeling fractions.

\begin{acknowledgments}
N.B. and G.G.  were supported in part by the US Department of Energy Grant
DE-FG03-91ER40662, Task C.  P.G. was  supported  in part by  the NFS
grant PHY-0756962 at the University of Utah. We would like to thank S. Nussinov and F. Avignone for
several important discussions about their work, and to   J. U Andersen, D. S.  Gemmell, D. V. Morgan, G. Hobler, and Kai Nordlund for some exchange of information.
\end{acknowledgments}

\appendix
\addappheadtotoc

\section{Penetration length of channeled ions}

Fig.~\ref{xMaxSiGe} shows the maximum distance, $x_{\rm max}(E)$ a channeled ion with initial energy $E$  propagates in a crystal channel, according to  the Lindhard-Scharff~\cite{Lindhard-Scharff, Dearnaley:1973} model of electronic energy loss, for a Si ion  channeled in a Si crystal and a Ge ion in a Ge crystal. This model is valid  for small enough energies,  $E < (M_1/2)Z_1^{4/3} v_0^2$ (where $v_0={e^2}/{\hbar} = 2.2 \times 10^8$ cm$/$sec is the Bohr's velocity~\cite{Lindhard:1965}. $M_1$ and $Z_1$ are the mass and charge of the propagating ion) which is $E< 24.3$ MeV for a Si ion propagating in a Si crystal and  $E< $ 188.7 MeV for a Ge ion propagating in a Ge crystal.
\FIGURE[h]{\epsfig{file=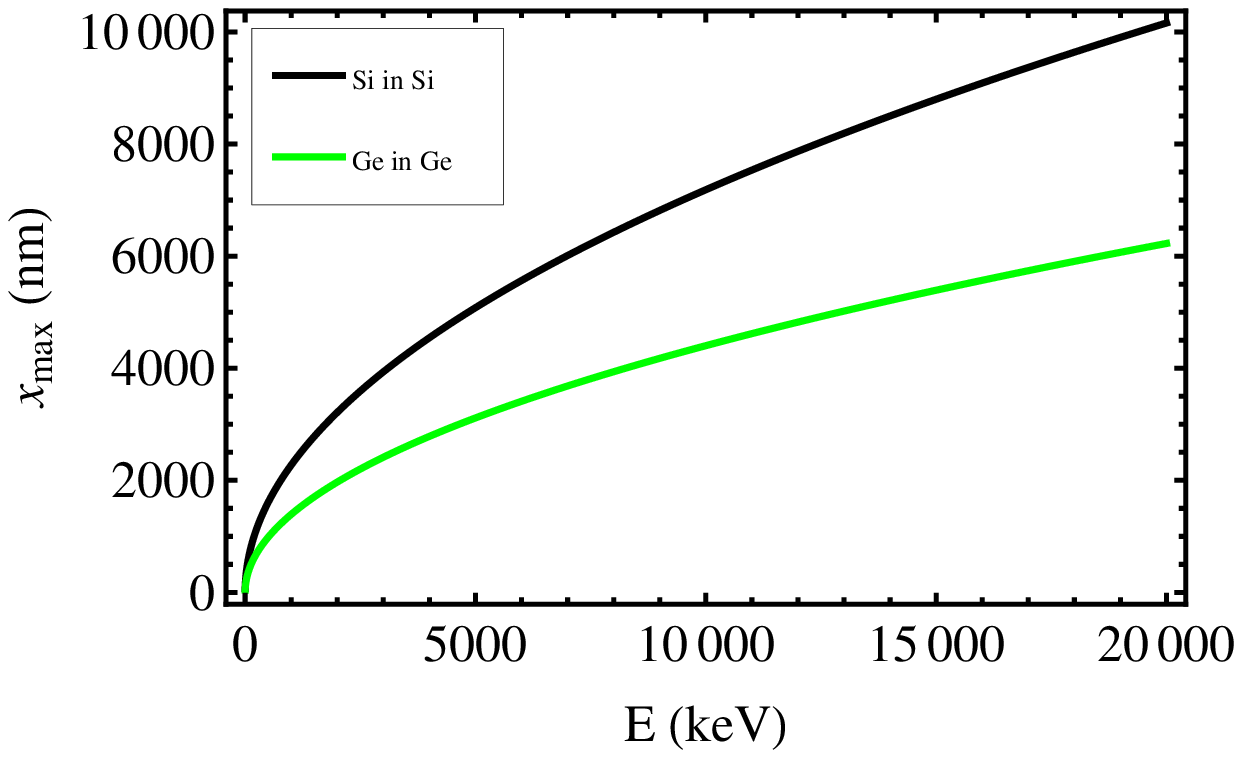,height=210pt}
      \caption{Maximum distance $x_{\rm max}(E)$ traveled  by channeled Si ions in Si (black) or Ge ions in Ge (green/gray).}%
	\label{xMaxSiGe}}
In this model the energy $E(x)$ as a function of the propagated distance $x$ and the initial energy $E$
 is the  solution of the following energy loss equation~\cite{Dearnaley:1973}
\begin{equation}
-\frac{dE}{dx}=Kv,
\end{equation}
where $v=\sqrt{2E/M_1}$ is the ion velocity and $K$ is the function
\begin{equation}
K=\frac{\xi_e 8 \pi e^2 N a_0 Z_1 Z_2}{ \left(Z_1^\frac{2}{3}+Z_2^\frac{2}{3}\right)^{\frac{3}{2}}v_0}.
\end{equation}
Here $\xi_e$ is a dimensionless constant of the order of $Z_1^{\frac{1}{6}}$~\cite{Dearnaley:1973}, $N$ is the number of atomic centers per unit volume, $a_0\simeq 0.53$ {\AA} is the Bohr radius of the hydrogen atom.
Explicitly, an ion with initial energy $E$ at $x=0$ has energy
\begin{equation}
E(x)=E\left(1-\frac{x}{x_{\rm max}}\right)^2
\label{Ex}
\end{equation}
 after traveling a distance $x$. The range of the propagating ion is
\begin{equation}
x_{\rm max}(E)=\frac{\sqrt{2M_1E_R}}{K}.
\label{xmax}
\end{equation}
 Fig.~\ref{xMaxSiGe} shows that even at energies of a few keV a channeled ion interacts with hundreds of lattice atoms. The characteristic interdistance of atoms along the channels is the lattice constant, i.e. approximately 0.5 nm for Si and Ge crystals (see Appendix B).

\section{Crystal structure of Si and Ge}

Silicon (Si) and  Germanium (Ge) crystals have a diamond cubic type lattice structure which consists of two interpenetrating face centered cubic (f.c.c.) lattices, displaced along the body diagonal of the cubic cell by one quarter of the  length of the diagonal. The unit cell, shown in Fig.~\ref{UnitCell}, has 8 atoms.  The lattice constant, the side of the cube in Fig.~\ref {UnitCell}, is $a_{\rm lat}=0.5431$ nm for Si and 0.5657 nm for Ge (from the Table 3.4 of Ref.~\cite{Appleton-Foti:1977}).
\FIGURE{\epsfig{file=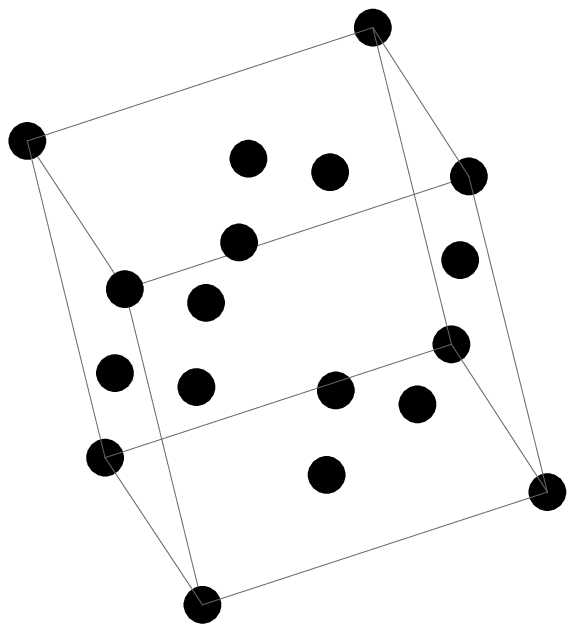,height=160pt}
\epsfig{file=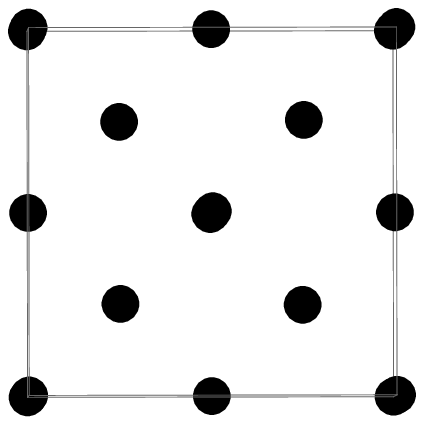,height=160pt}
      \vspace{-1.0cm}\caption{Unit cell of  a Si or Ge crystal (a) in three dimensions, and (b) projected on a plane. The black spheres represent Si or Ge atoms.}%
	\label{UnitCell}}

The atomic mass and atomic number of Si and Ge are $M_{Si}=28.09$ amu, $M_{Ge}=72.59$ amu, $Z_{Si}=14$ and $Z_{Ge}=32$. The Thomas-Fermi screening distances for two Si atoms and  two Ge atoms are $a_{SiSi}=0.4685 {\text {\AA} }  (Z_{Si}^{1/2} + Z_{Si}^{1/2})^{-2/3} =0.01225$ nm and $a_{GeGe}=0.4685 {\text {\AA} } (Z_{Ge}^{1/2} + Z_{Ge}^{1/2})^{-2/3} =0.009296$ nm respectively. Once an origin of the coordinate system is fixed on a lattice point $O$, any position vector of a point on the crystal lattice can be written as $\textbf{R}=n_1\textbf{a}+n_2\textbf{b}+n_3\textbf{c}$ with $n_1$, $n_2$, and $n_3$ specific integer numbers. The vectors $\textbf{a}$, $\textbf{b}$, and $\textbf{c}$ are the basis vectors of the crystal lattice, and are three noncoplanar vectors joining the lattice point $O$ to its near neighbors. For Si and Ge, the three vectors $\textbf{a}$, $\textbf{b}$, $\textbf{c}$ form a Cartesian frame and their length is $a_{\rm lat}/4$. The integers $n_1$, $n_2$, and $n_3$ can be positive, negative, or zero. The direction of a crystal axis pointing in the direction $\textbf{R}$ is specified by the triplet $[n_1 n_2 n_3]$ written in square brackets, when $n_1$, $n_2$, and $n_3$ are positive or zero. Note that if there is a common factor in the numbers $n_1$, $n_2$, $n_3$, this factor is removed. Moreover, negative integers are denoted with a bar over the number, e.g.\ $-1$ is denoted as $\bar{1}$ and the $-y$ axis is $[0\bar{1}0]$ direction. The plane perpendicular to the $[n_1 n_2 n_3]$ axis is denoted by $(n_1 n_2 n_3)$. For example, the plane perpendicular to the $[100]$ axis is denoted by $(100)$, and that perpendicular to $[101]$ by $(101)$. The integers $n_1$, $n_2$, and $n_3$ are called Miller indices.

 In a cubic crystal, because of the symmetry of the unit cell, the directions $[100]$, $[010]$,  $[001]$, $[\bar{1}00]$, $[0\bar{1}0]$ and $[00\bar{1}]$ are equivalent. All directions equivalent to the $[n_1 n_2 n_3]$ direction are denoted by $<$$n_1n_2n_3$$>$ in angular brackets. For example, $<$100$>$ indicates all the six directions mentioned.  Similarly, $<$211$>$ and $<$311$>$ indicate twelve different directions each.

  When the unit cell has cubic symmetry, we can indicate all planes that are equivalent to the plane $(hkl)$ by curly brackets $\{hkl\}$. For example, the indices \{100\} refer to the six planes $(100)$, $(010)$, $(001)$, $(\bar{1}00)$, $(0\bar{1}0)$, and $(00\bar{1})$. The negative sign over a number denotes that the plane cuts the axis on the negative side of the origin.  Similarly, $<$210$>$ and $<$310$>$ each indicate twelve different planes.

  Counting all the axes and planes we mentioned above, the total is 74.

The interatomic spacing $d$ in atomic rows and the interplanar spacing $d_{pch}$ (``pch" stands for ``planar channel") of atomic planes of monatomic diamond crystals, are obtained by multiplying the respective lattice constant by the following coefficients~\cite{Gemmell:1974ub}:
\begin{itemize}
  \item Rows: $<100>: 1$, $<110>: 1/\sqrt{2}$, $<111>: 3\sqrt{3}/4$, $<211>: \sqrt{6}/2$, $<311>: 3 \sqrt{11}/4$
  \item Planes: $\{100\}: 1/4$, $\{110\}: 1/2\sqrt{2}$, $\{111\}: \sqrt{3}/4$, $\{210\}:  1/(4 \sqrt{5})$, $\{310\}: 1/(2 \sqrt{10})$
\end{itemize}

The Debye temperatures for Si and Ge are $\Theta=490$ K and $\Theta=290$ K,
 respectively ~\cite{Gemmell:1974ub, Hobler}.

\section{Probability of correlated channels}

In this paper, as we did in Ref.~\cite{BGG-I}, we treat channeling along different channels as independent events when computing the probability $\chi_{\rm rec}(E,\hat{\bf q})$ in Eq.\ref{chirec} that an ion with initial energy $E$ and direction $\hat{\bf q}$  enters any of the available channels. Available channels are those whose axis or plane, respectively,  form and angle with the direction  $\hat{\bf q}$ smaller than the critical channeling angle for the particular channel.

 In Appendix D of Ref.~\cite{BGG-I} we showed that we can obtain an upper limit to the channeling probability of overlapping channels by replacing the intersection of the complements of the integration regions in Eqs.~\ref{chiaxial} and \ref{chiplanar} with the inscribed cylinder of radius $r_{\rm MIN}$  equal to the minimum of the $r_{i, {\rm min}}$ or $x_{i, {\rm min}}$ among the overlapping channels. We find that the two methods give practically indistinguishable results for Si and Ge (as we did in Ref.~\cite{BGG-I} for NaI), as clearly shown in
Fig.~\ref{FracSiGe-Max} for some particular  examples. Thus, the method we use is adequate for our purpose of providing upper bounds to the channeling fractions.
\FIGURE[h]{\epsfig{file=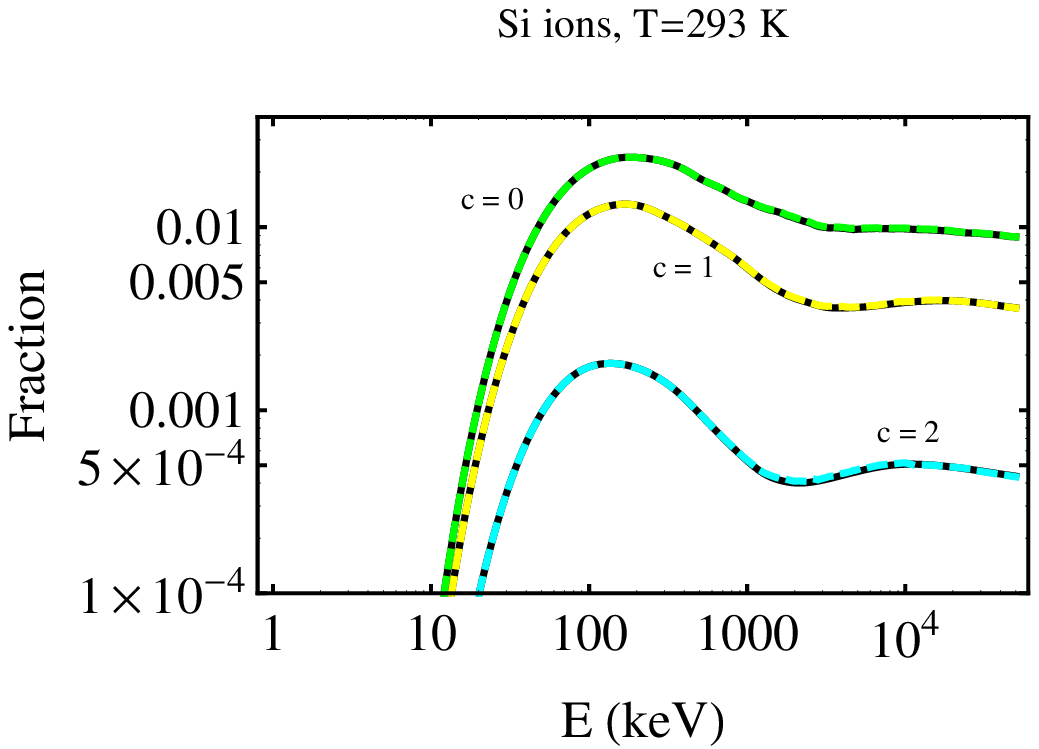,height=150pt}
\epsfig{file=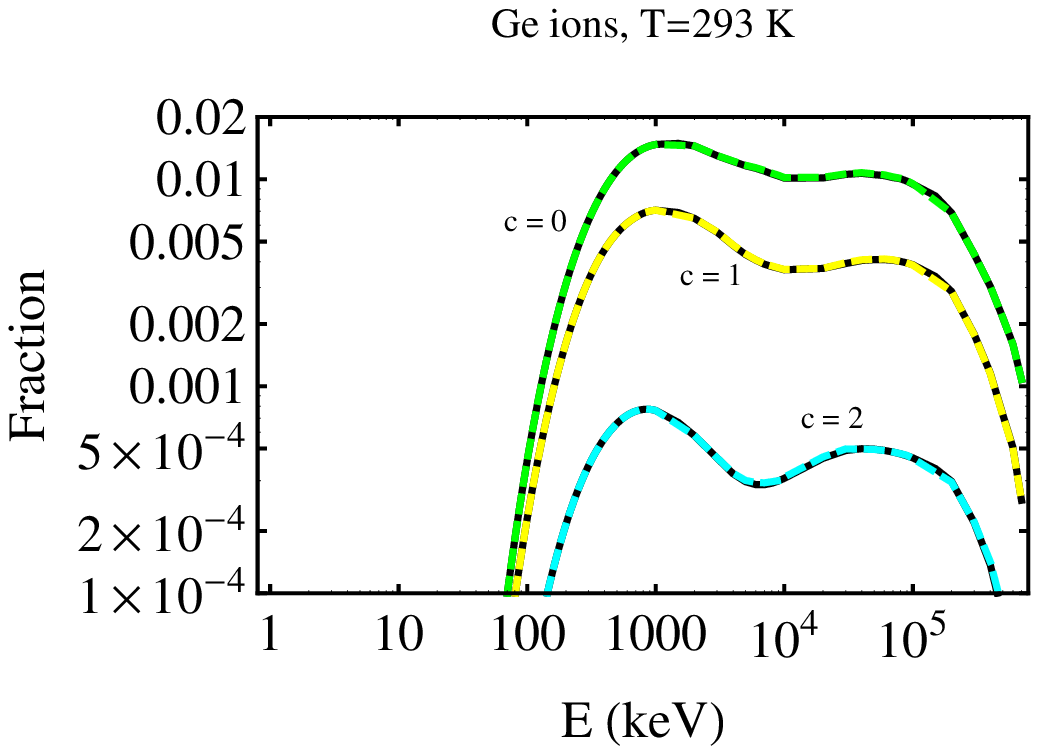,height=150pt}
        \caption{Maximum channeling fractions for $c_1=c_2=c$ and $c=0$ (dashed green), $c=1$ (dashed yellow) and $c=2$ (dashed cyan) compared with the results for the same models of our method of Section 3.2 (solid black lines) or (a) Si ions propagating in a Si crystal and (b) Ge ions propagating in a Ge crystal at $T=293$ K. Notice that the lines for the same model overlap.}%
	\label{FracSiGe-Max}}

Fig.~\ref{OneChannel-SiGe} shows the channeling fractions of Si ions propagating in a Si crystal and Ge ions propagating in a Ge crystal for individual channels with $c_1=c_2=1$ and T$=293$ K. The black and green (or gray) lines correspond to single axial and planar channels respectively.
\FIGURE[h]{\epsfig{file=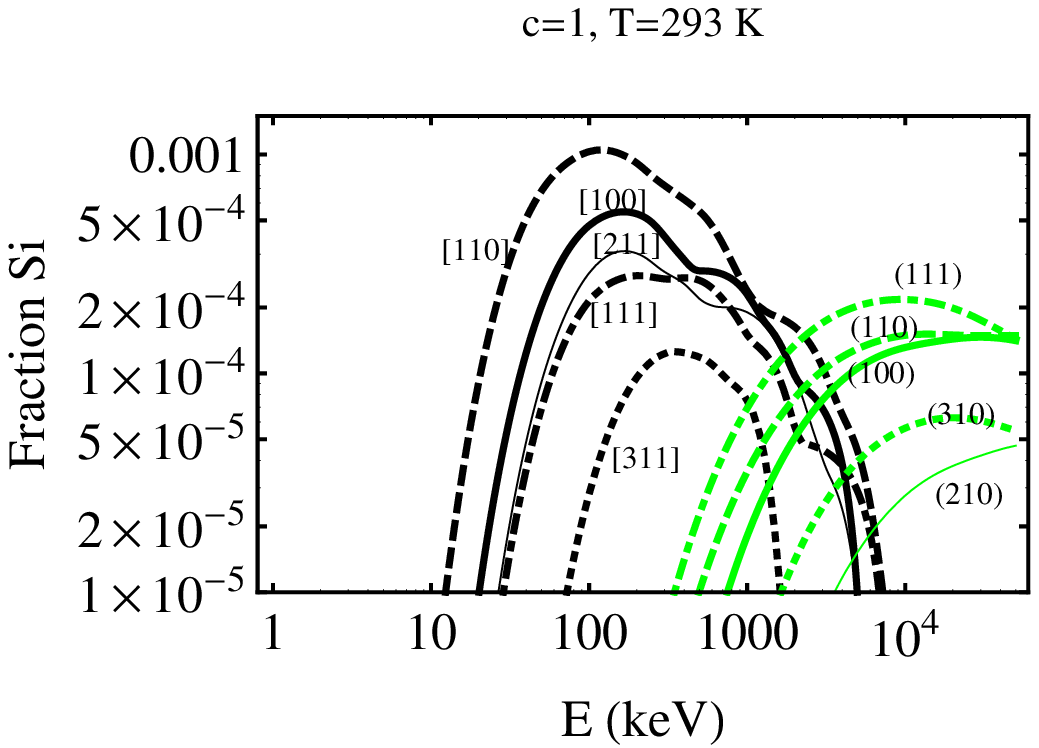,height=150pt}
\epsfig{file=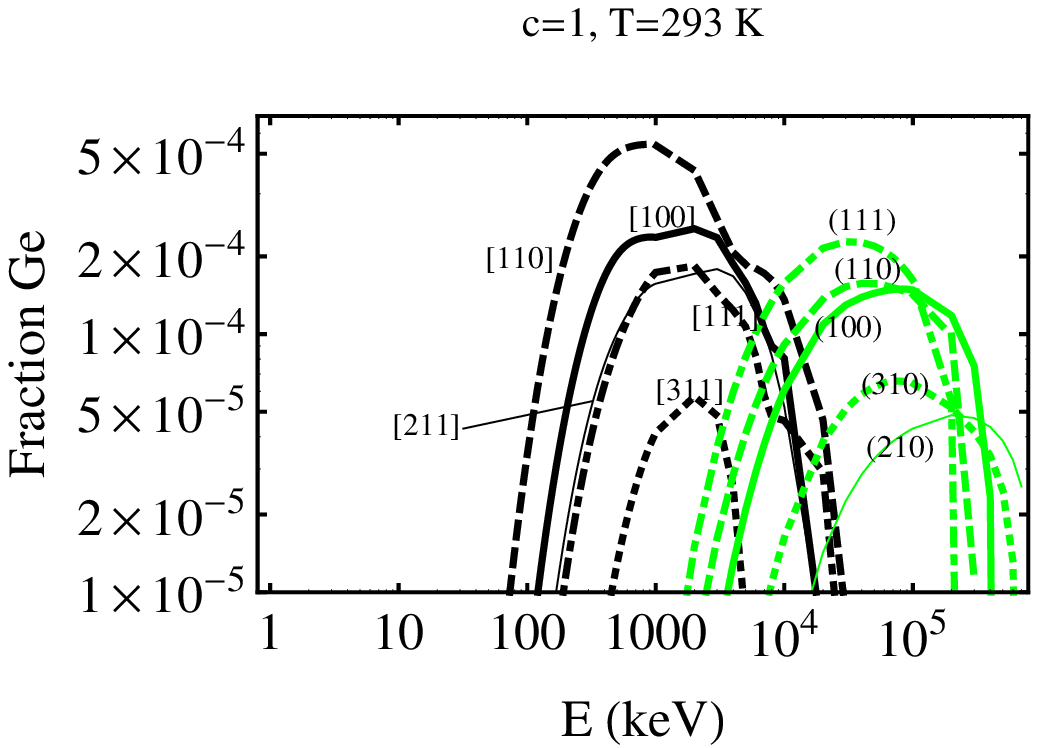,height=150pt}
        \caption{Channeling fractions of (a) Si ions propagating in a Si crystal and (b) Ge ions propagating in a Ge crystal for single planar (green/gray lines) and axial (black lines) channels, as function of the recoil energy $E$, for T$=293$ K and $c_1=c_2=1$.}%
\label{OneChannel-SiGe}}

Fig.~\ref{OneChannel-SiGe} shows that at low $E$ channeling is dominated by axial channels which do not overlap, so treating them as independent is strictly correct. However, at the transition energy of 1 to 10 MeV at which axial and planar channels are both equally important, and at higher energies at which planar channels dominate, the overlap of one axial and two or more planar channels, or the overlap of two or more planar channels among themselves, makes the channeling along them not necessarily uncorrelated. Still we find that considering channeling along different channels as independent is a good approximation if we are interested in providing upper bounds to the channeling fractions.

\end{document}